\documentclass[aps,prb,showpacs,twocolumn,amsmath,amssymb]{revtex4}
\usepackage{amssymb, amsmath}
\makeatother
\usepackage{graphicx}
\usepackage{float}
\usepackage{txfonts}
\usepackage{ulem}
\usepackage{bm}
\newcommand{\br}{{\bf r}}
\newcommand{\bfk}{{\bf k}}
\newcommand{\vk}{{\bf k}}
\newcommand{\vq}{{\bf q}}
\newcommand{\vp}{{\bf p}}
\newcommand{\vQ}{{\bf Q}}

\newcommand{\up}{\uparrow}
\newcommand{\dn}{\downarrow}

\usepackage{color}
\usepackage{ulem}
\begin{document}

\title{Broken-symmetry phases of  interacting nested  Weyl and Dirac  loops }

\author{
Miguel A. N. Ara\'ujo$^{1,2}$,  and Linhu Li$^3$
}

\affiliation{$^1$ CeFEMA, Instituto Superior
T\'ecnico, Universidade de Lisboa, Av. Rovisco Pais, 1049-001 Lisboa,  Portugal}

\affiliation{$^2$ Departamento de F\'{\i}sica,  Universidade de \'Evora, P-7000-671,
\'Evora, Portugal}

\affiliation{$^3$ Department of Physics, National University of Singapore, 117542, Singapore}

\begin{abstract}
We study interaction-induced
broken symmetry phases
 that can arise in metallic or semimetallic  band structures with two nested Weyl or Dirac loops. 
The odered phases can be
 of the charge or (pseudo)spin density wave type, or superconductivity from  interloop pairing.
A general analysis for two types of Weyl
 loops is given, according to whether a local reflection symmetry  in momentum  space  exists or not,
 for Hamiltonians  having  a global PT symmetry.
The resulting density-wave phases always  have lower total energy, and 
can be metallic, insulating, or semimetallic (with nodal loops), 
depending 
on both the reflection symmetry of the loops and the symmetry transformation that maps one loop onto the other.
We  extend this study to
nested $\mathbb{Z}_2$ nodal lines, for which the ordered phases
include also nodal point and nodal chain semimetals, and to spinful Dirac nodal lines. 
 Superconductivity from interloop pairing can be  fully gapped only if the initial double loop system is semimetallic.
\end{abstract}

\pacs{71.10Fd, 71.30.+h, 71.45.Lr,  75.30.Fv, 74.20.Rp}

\maketitle

\section{Introduction}

Since the discovery of topological insulators, band structures  of fermionic systems with non-trivial 
momentum-space topology have received much attention in modern condensed matter physics.  
Their low energy description  involves Dirac-like band dispersions, which in some cases imply
gapless 
 band structures characterized by the presence of nodal points or lines.
Among these, are the
  nodal line semimetals (NLSMs) \cite{mullen,hklattice}.
A NLSM has valence and conduction bands touching along one-dimensional (1D) lines in the three-dimensional (3D) momentum space, and feature two-dimensional (2D) ``drumhead" surface states surrounded by the nodal lines \cite{Burkov,Yang2014,Chen2015,Weng2015,Zhang2016}. 
Contrary to the well-studied topological insulating phases and nodal point semimetals, 
the 1D nodal lines of NLSMs provide rich topological structures such as links and knots \cite{Links_knots1,Links_knots2,Links_knots3,Links_knots4,Links_knots5}, which cannot be described unambiguously by a single sign (e.g. the $Z_2$ number) or a integer (e.g. the Chern number) \cite{multi_links_Li}. 
On the other hand, a variety of gapped and gapless topological phases have been predicted in NLSMs
(while possibly breaking certain symmetries). For example, a spin-orbit interaction can induce 3D Dirac semimetals from a NLSM \cite{SOC1,SOC2}, and periodical driving such as linear or circular polarized light may induce different types of nodal points \cite{light_driven1,light_driven2,light_driven3,light_driven4,light_driven5,light_driven6}. By introducing various types of extra gapped terms, a NLSM can also be driven into several different types of topological insulators, including the recently discovered high-order topological insulators \cite{TI_NL1,TI_NL2}.

Spontaneous symmetry breaking  from interactions 
  in three dimensional systems with Weyl/Dirac nodal points or lines have also been  addressed. 
For  single nodal loop (NL) systems, superconducting and charge (or spin) density wave instabilities have been 
 investigated using renormalization of fermionic interactions\cite{Nandkishore2}, including  also  the mean-field
description of the ordered phases\cite{Nandkishore1,roy,ryu}.

 Certain symmetries, such as spatial inversion or time-reversal, 
imply that  Weyl nodes must occur at an even number of Brillouin Zone (BZ) points.  
Charge and spin density waves, as well as superconducting phases, 
which arise from nested spherical Fermi surfaces (FSs) in  doped (or uncompensated)
 Weyl/Dirac  points have been discussed\cite{wangye}.
   Weyl or Dirac NLs, on the other hand,  do not necessarily have to exist in pairs. 
   Although two-loop semimetals 
have not yet been found in nature, pairs of linked NLs (or
    Hopf-link structures) have been theoretically proposed\cite{Links_knots3,Links_knots4,Links_knots6}. 
 Furthermore,  
a class of  NLs  protected by a combination of inversion
 and time-reversal has recently been discussed\cite{Z2_loop},  which 
 carry $\mathbb{Z}_2$ monopole charges, and  must therefore be created or annihilated  in pairs.

This has motivated us to address the  spontaneous symmetry breaking from  short range 
interactions  in two-loop band structures, when the NLs are related
through a nesting vector $\vQ$ in the BZ. We describe the density wave and superconducting phases, which
can be metallic, semimetallic (with double NLs) or fully gapped.    
A systematic analysis for two-band NL  models is given, where the NLs can either satisfy a local
reflection symmetry in the loop plane or not.  
If a global PT symmetry exists, then a symmetry operation can relate the two NLs. These properties combined
determine the nature of the ordered phases.  We also study specific four-band models that have 
appeared in the recent literature, such as the $\mathbb{Z}_2$ NLs, 
 and NLs arising from perturbed Dirac points.  
Superconducting phases arising from  pairing of fermions in different loops 
are also considered for all cases of singlet and triplet gap functions
in loop space as well as  (pseuso)spin space. But we have restricted our search to order parameters
with time-reversal symmetry (TRS) and fully gapped phases, because the latter are expected to be more stable. 
The possibility of gap functions with a winding number, which break TRS, is not addressed here.

The structure of the paper is as follows.
In Section \ref{modelsec} we introduce the local $k\cdot p$ Hamiltonians for two-band Weyl NLs, which can either  satisfy
a local reflection symmetry in the NL plane, or not. The Hubbard interaction and the
density wave order parameters associated with  the  NL nesting vector are also introduced.
 The density wave phases are described in Section  \ref{densec}, and 
Section \ref{examples}
 is devoted to example models and to a four-band system that was  not included in the general analysis of the
previous Sections: the nested $\mathbb{Z}_2$ NLs.
In Section \ref{Diracsec}, we study spin degenerate Dirac NLs    and also NLs arising from perturbed Dirac points.
   The superconducting pairing between nested NLs  is studied in Section \ref{supersec}. The analysis is 
 focused on interloop pairing and time-reversal symmetric order parameters.
 In Section  \ref{concsec} we present our conclusions.

\section{Model}
\label{modelsec}

We consider spinless fermions and 
let  $\boldsymbol\tau$ denote the Pauli matrices acting on the pseudo-spin (orbital) degree of freedom. 
We assume  that the band structure has two degenerate  loops.
 If the Hamiltonian has PT symmetry,  both loops involve the same Pauli matrices, $\tau_a\,,\tau_b$,
  so that each one can be locally described by  $k\cdot p$ Hamiltonians:
\begin{subequations}
 \begin{eqnarray}
H_0(\bfk)&=& v_1\left( p_{\parallel} - p_o \right) \tau_a + v_2 p_\perp \tau_b \,, \\
H_0(\bfk+\ \vQ)&=& g_1v_1\left( p_{\parallel} - p_o \right) \tau_a +g_2 v_2 p_\perp \tau_b\,,
\end{eqnarray}
\label{loop1}
\end{subequations}
Here, and throughout the paper, $\bf p=\hbar\vk$ and the subscripts $\perp$($\parallel$) refer to the components perpendicular (parallel) to
the loop plane, and 
$g_{1(2)}$ are  + or - signs. 
 The loops are nested by the  vector ${\vQ}$. 
We shall refer to the NLs in Eq. (\ref{loop1}) as ``model-1'' loops.
Such NLs
 are protected by a local reflection\cite{bian} in the loop plane, ${\cal R}=(p_\perp\rightarrow -p_\perp)\otimes \tau_a$. 
Such a NL can be topologically characterized by a $\pi$ Berry phase along a trajectory enclosing the NL \cite{Burkov,Zhang2016}.
At zero chemical potential,   the system is a semimetal and the  FS
consists of the two nested NLs.
 We shall  also take non-degeneracy into account by considering an energy offset $\delta$ between
the loops and make the replacement 
$H_0(\bfk)\rightarrow H_0(\bfk) -\delta$,
$H_0(\bfk+\ \vQ)\rightarrow H_0(\bfk+\ \vQ) +\delta$.
For positive $\delta$ and  zero chemical potential, the FSs are torus-shaped, the one from $H_0(\bfk+\ \vQ) $   
is in the lower (hole)  band, while the FS from $H_0(\bfk)$ is in the upper (electron) band.

However, NLs are not necessarily protected by  reflection symmetry. Here we also
consider a more general model
of nested NLs without reflection symmetry. The Hamiltonian reads
\begin{subequations}
 \begin{eqnarray}
H_0(\bfk)&=& \left[v_{1\parallel}\left( p_{\parallel} - p_o \right) + v_{1\perp} p_\perp \right] \tau_a + v_2 p_\perp \tau_b\,,
 \\
H_0(\bfk+\ \vQ)&=& \left[g_1'v_{1\parallel}\left( p_{\parallel} - p_o\right)+  g_2'v_{1\perp}p_\perp \right]\tau_a +
g_2 v_2 p_\perp \tau_b\,,
\nonumber\\
\end{eqnarray}
\label{loop2}
\end{subequations}
where  $g_{1(2)}'$ are  + or - signs. 
We refer to these as ``model-2'' loops.
The extra $p_{\parallel}$ in the $\tau_a$ term changes the pseudospin texture near a NL,
 but does not affect the topological properties associated with the Berry phase.
 Examples of both types of NLs will be given in 
Section  \ref{modelsec}.
  The above  two types of loops respond differently to the interactions, as shown  in the following sections.

Normaly, one should expect that a perturbation
arises that will lift the degeneracy between nested FSs. 
The perturbation may result from interactions and, in a normal system, usually 
takes the form  of  some charge or spin wave with the wave vector  ${\vQ}$.
Also,  superconducting pairing between fermions in different NLs will be considered.

In the rest of the paper we shall  set  to unity
the velocity prefactors in the Hamiltonians (\ref{loop1}) and (\ref{loop2}), 
as they are not really necessary for the analysis that follows.

\section{Density wave phases}
\label{densec}
\subsection{Interaction and mean-field theory}
We introduce a Hubbard interaction,
\begin{eqnarray}
\hat U &=& U\sum_\br \hat n_1(\br)\hat n_2(\br)\,,
\label{localU}
\end{eqnarray}
where the indices $1,2$ refer to the orbital degree of freedom.
Doing a mean field theory decoupling, the interaction takes the form
\begin{eqnarray}
\hat U_{MF} = U \sum_\br \left[
 \langle n_1(\br) \rangle \hat n_2(\br)  +  \hat n_1(\br)\langle n_2(\br)\rangle
- \langle n_1(\br) \rangle \langle n_2(\br)\rangle  \right]\,.
\label{mfloop}
\end{eqnarray}
A pseudospin density wave (PSDW) phase  with the same nesting wave vector  ${\vQ}$ 
is characterized by
 \begin{eqnarray}
  \langle n_j(\br)\rangle  = \frac 1 2 n + \bar m (-1)^j\cos({\vQ}\cdot\br)\,,
\end{eqnarray}
where $\bar m$ is the amplitude and $j=1,2$. 
Although this type of ordering  describes an imbalance in  orbital occupation,  it
is not a charge density wave (CDW) because the charge at site $\br$ is spatially constant, $n$. \
Omitting the factor $(-1)^j$, then a true CDW is obtained.
We introduce the annihilation operator $\hat\psi_j(\br) $, 
 at point $\br$, with pseudospin index $j$.
Then, Eq. (\ref{mfloop}) can be rewritten as:
 \begin{eqnarray}
&&\hat U_{eff} =  \frac{Un}{2} \sum_\br \left( \hat\psi_1^\dagger(\br)  \hat\psi_2^\dagger(\br)    \right) \tau_0
 \left( \begin{array}{c} \hat\psi_1(\br) \\ \hat\psi_2(\br)  \end{array}  \right)\nonumber\\ &&+\
U \bar m \sum_\br \cos({\boldsymbol Q}\cdot\br) \left( \hat\psi_1^\dagger(\br)  \hat\psi_2^\dagger(\br)    \right) \tau_3
\left( \begin{array}{c} \hat\psi_1(\br) \\ \hat\psi_2(\br)  \end{array}  \right)\nonumber\\
&=&
 \frac{Un}{2} \sum_\bfk \left( \hat c_{\vk ,1}^\dagger \hat c_{\vk ,2}^\dagger    \right) \tau_0
  \left( \begin{array}{c} \hat c_{\vk ,1}\\ \hat c_{\vk ,2}  \end{array}  \right)
  \nonumber\\ 
&+&
\frac{U \bar m}{2} \sum_\bfk     \left[ 
 \left( \hat c_{\vk+\vQ,1}^\dagger  \hat c_{\vk+\vQ,2}^\dagger   \right) \tau_3
  \left( \begin{array}{c} \hat c_{\vk ,1}\\ \hat c_{\vk ,2} \end{array}  \right) + 
   \left( \hat c_{\vk,1}^\dagger  \hat c_{\vk,2}^\dagger   \right) \tau_3
  \left( \begin{array}{c} \hat c_{\vk +\vQ,1}\\ \hat c_{\vk+\vQ ,2} \end{array}  \right)\right]\nonumber\\
  \label{UEff}
\end{eqnarray}
Replacing $\tau_3\rightarrow\tau_0$ in Eq. (\ref{UEff}),  we can describe a true CDW. 

We write the effective Hamiltonian  $H_{eff}(\vk)$ matrix
in operator basis $( \hat c_{\vk ,1}, \hat c_{\vk ,2} \hat c_{\vk+\vQ,1}, \hat c_{\vk+\vQ,2})
\equiv( \boldsymbol c_{\bfk} \ \boldsymbol c_{\bfk+\vQ} )$
and introduce a factor  $\frac 1 2$ to avoid double counting of momenta in the BZ:
\begin{eqnarray}
\hat H_{eff}&=& 
\frac 1 2 
\sum_\vk \left(  \boldsymbol c^\dagger_{\bfk}\ \boldsymbol c^\dagger_{\bfk+\vQ}  \right)
H_{eff}(\vk) 
\left( \begin{array}{c}   \boldsymbol c_{\bfk} \\ \boldsymbol c_{\bfk+\vQ}   \end{array}\right)\,,
\\
H_{eff}(\vk) &=&  \left(  \begin{array}{cc}
H_0(\vk) & U\bar m \tau_\alpha\\
 U\bar m \tau_\alpha & H_0(\vk+\vQ)  
\end{array}\right)    
\,,
\label{martizHloops} 
\end{eqnarray}
where $\alpha=0$ for CDW, or $\alpha=3$ for PSDW.
The mean field equations for this Hamiltonian are derived in Appendix \ref{APMF}.

The effective Hamiltonian 
Eq. (\ref{martizHloops}) is by no means restricted to the case of a local interaction 
as in Eq.  
(\ref{localU}). A nearest-neighbor interaction, for instance, would produce an effective
Hamiltonian of the same form, but where the bare interaction parameter would be multiplied by
a $\vQ$-dependent form factor which could still be denoted by ``$U$''.
The nesting property of the Fermi surface leads to a divergence
of the susceptibilities in momentum space at the nesting wavevector\cite{gruner}. 
This always
leads to a density wave with momentum $\bf Q$, described by the mean field couplings
$<\hat c_{\vk}^\dagger\hat c_{\vk+\vQ}>$,
and the relevant interaction "$U$" would be the Fourier component of the interaction
for the nesting wavevector.

\subsection{model-1 loops}
\label{model1_general}

We now  introduce the Pauli matrices   $t_\mu$ operating   in loop space $(\vk,\vk+\vQ)$. 
For  type-1 models 
the unperturbed double loop Hamiltonian has the form 
 \begin{eqnarray}
H_{eff}^0(\vk) = \left( p_{\parallel} - p_o \right) t_i\tau_a+  p_\perp t_j\tau_b -\delta\ t_3
\label{H0eff}
\end{eqnarray}
where $i,j$ can 
only take values 0 or 3.  
The effective Hamiltonian  (\ref{martizHloops}) can be written as
\begin{equation}
H_{eff}(\vk)= H_{eff}^0(\vk) + U\bar m\ t_1 \tau_\alpha
\label{umgap}
\end{equation}
Supose that $\delta=0$, that is, degenerate loops at perfect compensation. 
It is clear that if  the perturbing term $U\bar m t_1 \tau_\alpha$
 anticommutes with only one  term of $H_{eff}^0(\vk) $  then the resulting system still is a double NL semimetal.
 On the other hand, if  $U\bar mt_1\tau_\alpha$
 anticommutes with $H_{eff}^0(\vk)$ then the resulting system is a gapped insulator, 
  and if $\left[U\bar mt_1 \tau_\alpha, H_{eff}^0(\vk)\right]=0$ then the original loops are shifted
  and a metallic phase arises with torus-shaped FSs, one of them hole-like, and the other electron-like.

Next we establish a criterion based on how a unitary transformation maps one loop onto the other.
 If the Hamiltonian has PT symmetry, 
one can always find a rotation through a Pauli matrix $\tau_\beta$
that maps  one model-1 loop at $\bfk$  into the other at $\bfk+\vQ$:
 \begin{eqnarray}
 H_0(\bfk)  =   \tau_{\beta} H_0(\bfk+\ \vQ) \tau_{\beta}   \,.
 \label{rotj}
 \end{eqnarray} 
It is then convenient to apply a unitary transformation to  the effective Hamiltonian in Eq. (\ref{martizHloops})
according to
 \begin{eqnarray}
A H_{eff}(\vk)  A^\dagger &=& \left( \begin{array}{cc}  
H_0(\bfk) -\delta & U\bar m \tau_\alpha\tau_{\beta}\\ U \bar m \tau_{\beta}\tau_\alpha &  H_0(\bfk) +\delta
 \end{array}   \right) \label{rotatedloops}\\
 A&=&\left( \begin{array}{cc}  1 & 0 \\ 0 &\tau_{\beta}  \end{array}   \right)\,.\label{Amatrix}
 \end{eqnarray}

The energy spectrum can be obtained by performing appropriate rotations on the matrix (\ref{rotatedloops}),
as shown in Appendix \ref{appa}. 
We list all the four possibilities as follows.
\begin{enumerate}
\item
 If $\tau_\alpha \tau_{\beta}=1$, the spectrum reads:
\begin{eqnarray}
E=\pm \sqrt{ \left( p_{\parallel} - p_o \right)^2+p_\perp^2}
\pm   \sqrt{U^2\bar m^2 + \delta^2}
 \,,\label{CDWk0}
\end{eqnarray}
(uncorrelated $\pm$ signs). In this case,  $H_0(\bfk)=\tau_\alpha H_0(\bfk+\vQ)\tau_\alpha$.
The density wave produces 
 a ``level repulsion'' effect by introducing (increasing) 
 an energy splitting between the degenerate (non-degenerate) NLs.
 The density wave phase  has two toroidal FSs, one hole-like and one electron-like.
 
\item
In the case where $\tau_\alpha \tau_{\beta}\propto\tau_a$, the energy spectrum is:
\begin{eqnarray}
E^2 &=& \left( p_{\parallel} - p_o \right)^2 + p_\perp^2+ U^2\bar m^2 +\delta^2 \nonumber\\
&\pm& 2\sqrt{\left( p_{\parallel} - p_o \right)^2(U^2\bar m^2 + \delta^2)+
 p_\perp^2\delta^2}\,,\label{CDWka}
\end{eqnarray}
which yields two NLs at $p_\perp=0$, 
$p_{\parallel}=p_o\pm\sqrt{U^2\bar m^2 + \delta^2}$. As $p_\parallel$ can only take a 
positive
value, the loop with the 
minus sign will shrink into a point and vanish when $\sqrt{U^2\bar m^2 + \delta^2}$ becomes larger then $p_o$.
\item
For $\tau_\alpha \tau_{\beta}\propto\tau_b$ we get:
\begin{eqnarray}
E^2 &=& \left( p_{\parallel} - p_o \right)^2 + p_\perp^2+ U^2\bar m^2 +\delta^2 \nonumber\\
&\pm& 2\sqrt{  p_\perp^2(U^2\bar m^2 + \delta^2)+
\left( p_{\parallel} - p_o \right)^2\delta^2}\,.\label{CDWkb}
\end{eqnarray}
This spectrum also gives two NLs, given by $p_\parallel = p_o$,
$p_\perp=\pm \sqrt{U^2\bar m^2 + \delta^2}$. Unlike  the previous case, these two loops only move along 
the $p_\perp$ direction when tuning $U$ or $\delta$, while their radii remain unchanged. 
\item
For the case $\tau_\alpha \tau_{\beta}\propto\tau_{c(\neq a,b)}$ we have
\begin{eqnarray}
E^2 &=& U^2\bar m^2 + 
 \left[ \sqrt{ \left( p_{\parallel} - p_o \right)^2+p_\perp^2} \pm \delta \right]^2 \,,\label{CDWkc}
\end{eqnarray}
which is then fully gapped. This is the case where $H_0(\bfk)=-\tau_\alpha H_0(\bfk+\vQ)\tau_\alpha$.  
\end{enumerate}

At half filling (zero chemical potential), all the above energy dispersions  lead to a lowering of the energy
for $U\bar m\neq 0$, so,  the density wave phase is energetically favorable. 
In a single NL, the density of states vanishes linearly at the chemical potential, so  the broken symmetry
phase  appears only for  $U$ above a  finite critical value, $U_{cr}$. 
In any case if the phase  transition is second order, then 
 $U\rightarrow U_{cr}^+ \Rightarrow\ \bar m\rightarrow 0$. 
 The PSDW cases occur for $U>U_{cr}>0$ but the CDW ordering requires negative $U<U_{cr}<0$, hence an attractive 
 interaction (see Appendices \ref{APMF}  and  \ref{MFloop}).

From Eq. (\ref{rotj}) we see that  $\tau_\alpha H_0(\bfk) \tau_\alpha =  
 \tau_\alpha\tau_{\beta} H_0(\bfk+\ \vQ) \tau_{\beta}\tau_\alpha$
and therefore, the density-wave  phase is a nodal line semi-metal if
\begin{equation}
H_0(\bfk)  \neq  \pm \tau_\alpha H_0(\bfk+\ \vQ)\tau_\alpha \,,\label{critCDW}
\end{equation}
where $\alpha=0$ for CDW, or $\alpha=3$ for PSDW.

For model-1 loops we can make the following observations regarding  symmetry. 
The Hamiltonian (\ref{rotatedloops}) for $\delta=0$ is chiral as it anti-commutes with the operator $\tau_{c\neq a,b}$, 
and contains the degenerate NLs.
This chiral symmetry can be broken by a term of the form:
 {\it (i)} $t_\mu\tau_c$ which may  fully gap the spectrum,
 or yield a  semimetal, depending on its exact form; 
 {\it (ii)} $t_\mu\tau_0$ which shifts the original loops and  leads to a metallic spectrum.
On the other hand, a term of the form $t_j\tau_a$ or $t_j\tau_b$ ($j=1,2,3$), preserves the chiral symmetry
and yields the NL semimetal even if $\delta t_3 \neq 0$ is already present, as shown in  Appendix \ref{appa}.

\subsection{model-2 loops}\label{model2_general}

As we shall see, the criteria (\ref{critCDW})  do not always apply to  model-2 loops.
 The type-2 loop Hamiltonian at $\mathbf{k}$ is  (omitting velocity prefactors):
 \begin{eqnarray}
H_0(\bfk)&=& \left( p_{\parallel} - p_o  + p_\perp\right)\tau_a + p_\perp \tau_b\,,
\end{eqnarray}
and the loop at  $\bfk+ \vQ$ can always be related to that in $\bfk$ by either:
{\it (i)} a rotation through a Pauli matrix if $g'_1g'_2=1$ in Eq. (\ref{loop2}); or  {\it (ii) } a reflection in the loop plane
if $g'_1g_2'=-1$.
If {\it (i)} holds, then one can again 
rotate the   effective Hamiltonian according to equations  (\ref{rotatedloops})-(\ref{Amatrix}), and
the resulting spectra can be obtained from
 Eqs (\ref{CDWk0})-(\ref{CDWkc}) for model-1 loops,
 with the replacement
$p_{\parallel} - p_o\rightarrow p_{\parallel} - p_o+p_\perp$. 
 But in case {\it (ii)}  the two NLs can be related through a reflection in the NL's plane,
 \begin{subequations}
\begin{eqnarray}
H_0(\bfk) &=&  {\cal R} H_0(\bfk+\ \vQ)  {\cal R}^\dagger\,,\\
{\cal R} &=&  (p_\perp\rightarrow - p_\perp) \tau_{\beta}\,, 
\end{eqnarray}\label{reflection}
\end{subequations}
which  corresponds to the following cases, depending on $\tau_{\beta}$:
\begin{eqnarray}
\begin{aligned}
& H_0(\bfk+\ \vQ)=\\
&= \left( p_{\parallel} - p_o  - p_\perp\right)\tau_a - p_\perp \tau_b\equiv H_0(-k_\perp) &\beta&=0;\\
&= \left( p_{\parallel} - p_o  - p_\perp\right)\tau_a + p_\perp \tau_b &\beta&=a;\\
&= \left[ -\left( p_{\parallel} - p_o\right)  + p_\perp\right]\tau_a - p_\perp \tau_b & \beta&=b;\\
&= \left[ -\left( p_{\parallel} - p_o\right)  + p_\perp\right]\tau_a + p_\perp \tau_b & \beta&=c\neq a,b\,.
\end{aligned}
\end{eqnarray}
We apply the same rotation to the effective Hamiltonian,
 using Eqs. (\ref{rotatedloops})-(\ref{Amatrix}):
 \begin{eqnarray}
A H_{eff}(\vk)  A^\dagger &=& \left( \begin{array}{cc}  
H_0(\bfk) & U\bar m \tau_\alpha\tau_{\beta}  \\ U \bar m \tau_{\beta}\tau_\alpha &  H_0(-k_\perp)
 \end{array}   \right)\label{difficult}\,.
 \end{eqnarray}

\begin{figure}
\includegraphics[width=1\linewidth]{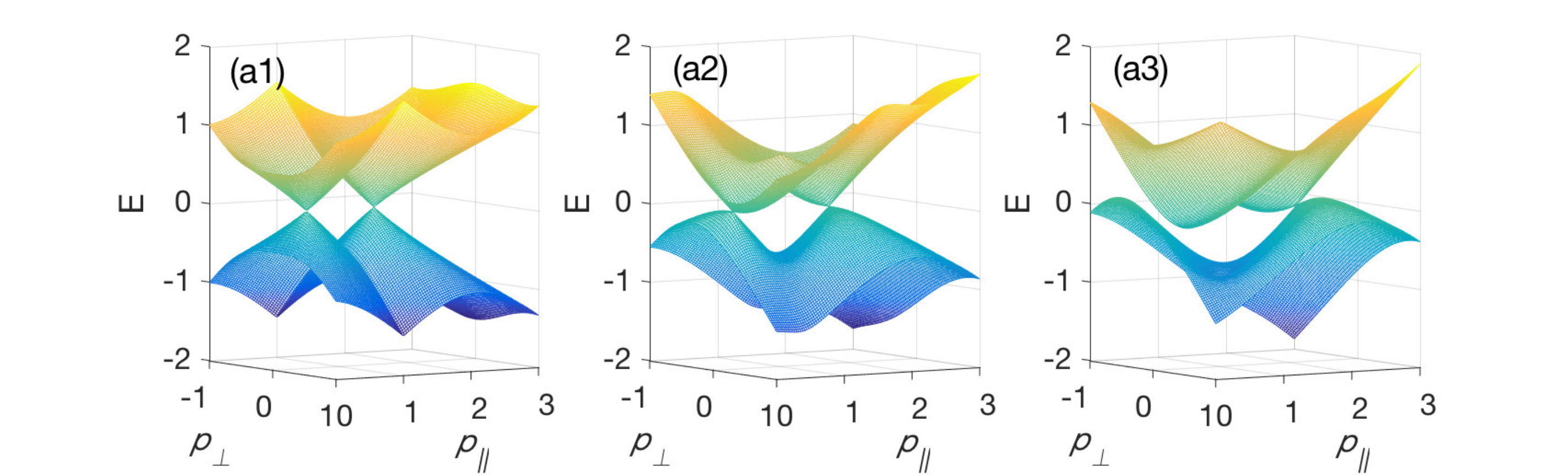}
\includegraphics[width=1\linewidth]{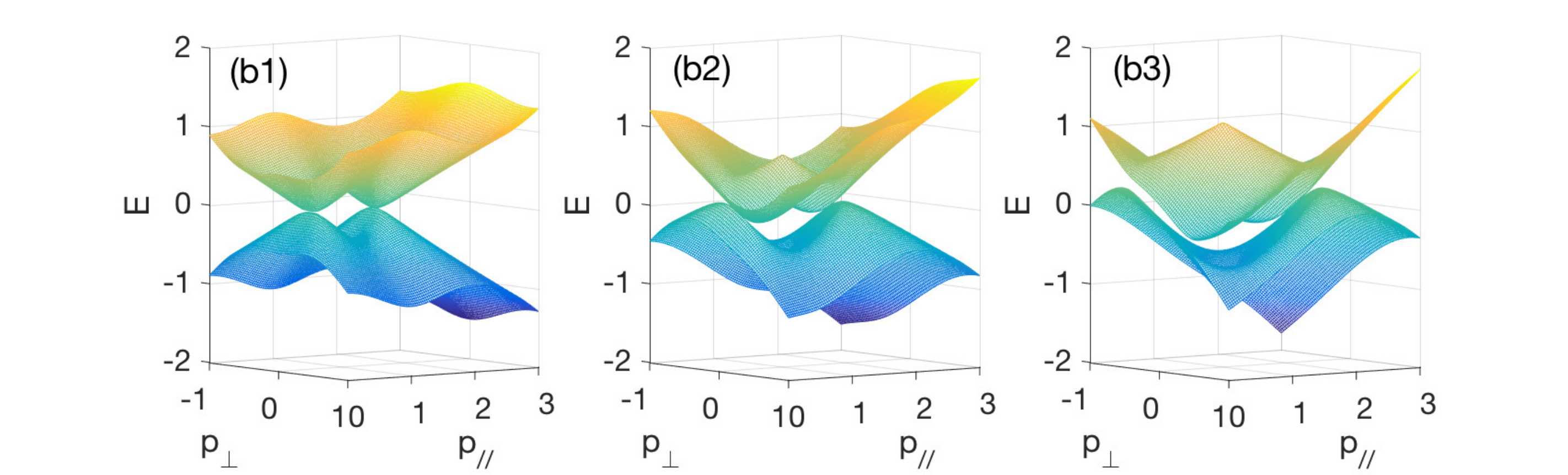}
\includegraphics[width=1\linewidth]{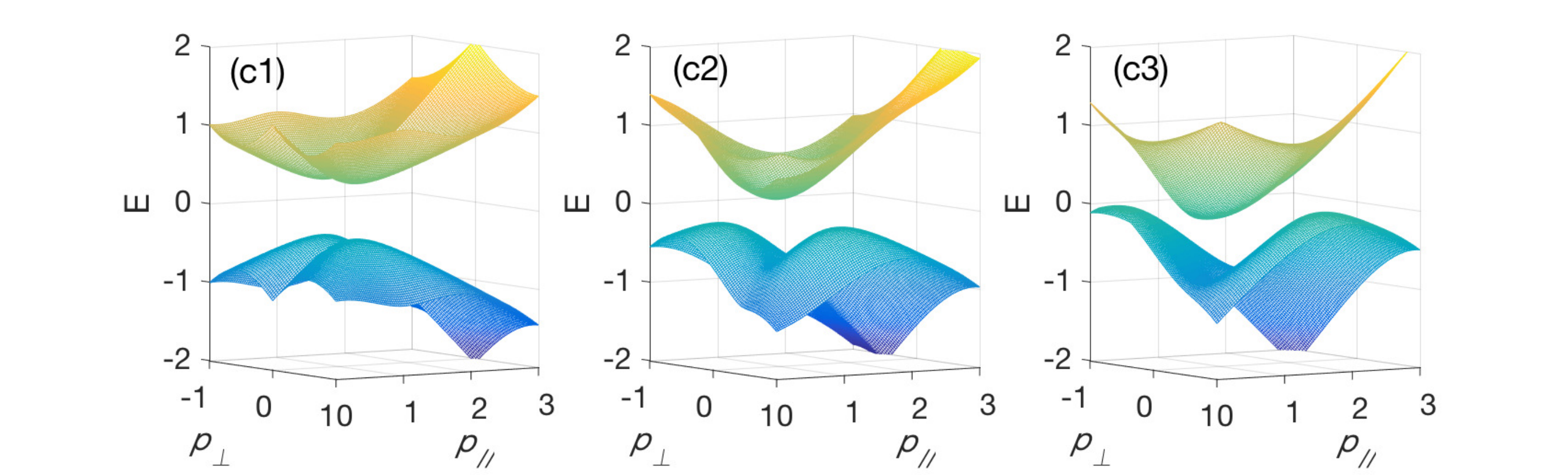}
\includegraphics[width=1\linewidth]{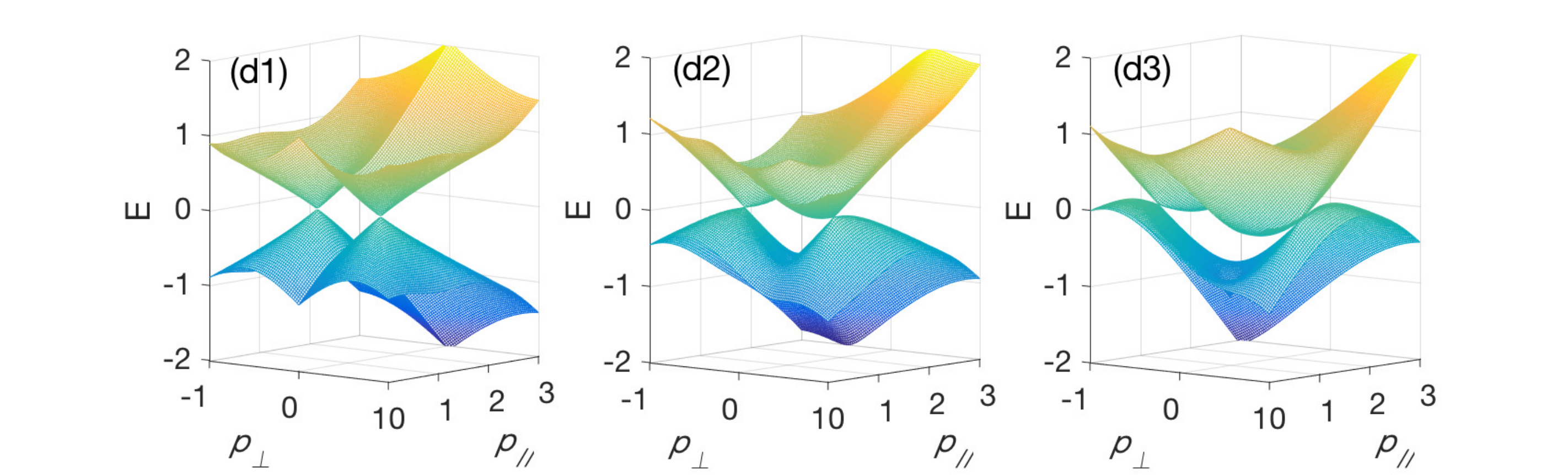}
\caption{The energy dispersion of the two inner bands of the spectrum. 
Panels (a)-(d) correspond to the four cases  in Eqs. (\ref{mode2_case1}), (\ref{mode2_case2}), (\ref{mode2_case3}), and (\ref{mode2_case4}) respectively. The  energy offset $\delta$ is: (a1)-(d1) $\delta=0$; (a2)-(d2) $\delta=0.5$; and (a3)-(d3) $\delta=1$. The other parameters are: $p_o=1$, and  $U\bar m =0.5$. }\label{fig1}
\end{figure}

For finite $\delta$ one cannot write the energy dispersion in closed form. We analytically deal
with the  degenerate case at perfect compensation,  $\delta=0$, below, 
and show  numerical results for nonzero $\delta$ in Fig.\ref{fig1}. 
The figure  shows the two inner bands
 of Hamiltonians (\ref{mode2_case1}), (\ref{mode2_case2}), (\ref{mode2_case3}), and (\ref{mode2_case4}) in the two-dimensional space of $(p_{\parallel} ,p_\perp)$. A NL for $U\bar m = \delta=0$
then looks like a Dirac cone  at point  $(p_o ,0)$. The splitting of the original NLs can be seen as the appearance of 
two Dirac cones in the plot. For finite $\delta$, the Dirac cone axis is tilted.

Similarly to the discussion for model-1 loops, we also list all the four possibilities.
\begin{enumerate}
\item
If $\tau_\alpha\tau_{\beta}=1$:
\begin{eqnarray}
A H_{eff}(\vk)  A^\dagger &=& 
\left( p_{\parallel} - p_o  \right)\tau_a +  p_\perp t_3\tau_a +  p_\perp t_3\tau_b\nonumber\\
&+& U\bar m \ t_1 - \delta t_3\,,\label{mode2_case1}
 \end{eqnarray}
For $\delta=0$ (perfect compensation) the spectrum obeys
 \begin{eqnarray}
E^2 = p_\perp^2 + \left[ \sqrt{p_\perp^2 + (U\bar m)^2} \pm \left(  p_{\parallel} - p_o\right)
\right]^2
 \end{eqnarray}
which has two nodal lines for $p_\perp=0$ and $p_{\parallel} - p_o=\pm U\bar m$.  
 By turning on $\delta$, the two NLs are tilted along the $p_{\perp}$ direction, and move 
 along the $p_{\parallel}$ direction, as shown in Fig. \ref{fig1}(a). It can also be seen  from Eq. (\ref{mode2_case1})
that if $p_\perp=0$ the dispersion relation has  two nodal lines: $|p_{\parallel} - p_o|=\sqrt{U^2\bar m^2+\delta^2}$. 
Therefore, one of the loops will shrink into the origin as $p_{\parallel}\rightarrow 0$ when 
 $\delta^2+U^2\overline{m}^2\rightarrow p_0^2$, 
and become gapped for larger $\delta$. 
\item 
If $\tau_\alpha\tau_{\beta}\propto\tau_a$:
\begin{eqnarray}
A H_{eff}(\vk)  A^\dagger &=& 
\left( p_{\parallel} - p_o  \right)\tau_a +  p_\perp t_3\tau_a +  p_\perp t_3\tau_b\nonumber\\
&+& \epsilon_{k\alpha a}U\bar m \ t_2\tau_a - \delta t_3\,,\label{mode2_case2}
 \end{eqnarray}
For $\delta=0$ (perfect compensation) the spectrum obeys
 \begin{eqnarray}
E^2 &=&   \left(  p_{\parallel} - p_o\right)^2 + 2 p_\perp^2 +(U\bar m)^2 \nonumber\\
& \pm &
2 \sqrt{ 
\left(  p_{\parallel} - p_o\right)^2 \left(   p_\perp^2 +U^2\bar m^2\right)  + p_\perp^2 U^2\bar m^2
}
 \end{eqnarray}
which, for $\delta=0$, has the same two nodal 
lines as in the previous case, and behavior of these lines
 with nonzero $\delta$ is also identical. However, these loops show a
 quadratic dispersion along the $p_\perp$ direction, as shown in Fig. \ref{fig1}(b).
\item 
If $\tau_\alpha\tau_{\beta}\propto\tau_b$:
\begin{eqnarray}
A H_{eff}(\vk)  A^\dagger &=& 
\left( p_{\parallel} - p_o  \right)\tau_a +  p_\perp t_3\tau_a +  p_\perp t_3\tau_b\nonumber\\
&+& \epsilon_{k\alpha b}U\bar m \ t_2\tau_b - \delta t_3\,, \label{mode2_case3}\\
E^2 &=& p_\perp^2 + \left(
 \sqrt{  \left(  p_{\parallel} - p_o\right)^2 + U^2\bar m^2} \pm  p_\perp
 \right)^2
 \end{eqnarray}
which is a fully gapped insulator, and remains gapped with nonzero $\delta$ [Fig. \ref{fig1}(c)].
\item 
If $\tau_\alpha\tau_{\beta}\propto\tau_c(\neq a,b)$:
\begin{eqnarray}
A H_{eff}(\vk)  A^\dagger &=& 
\left( p_{\parallel} - p_o  \right)\tau_a +  p_\perp t_3\tau_a +  p_\perp t_3\tau_b\nonumber\\
&+& \epsilon_{k\alpha c}U\bar m \ t_2\tau_c - \delta t_3\,,\label{mode2_case4}
\end{eqnarray}
which for $\delta=0$ has the spectrum
\begin{eqnarray}
E^2 &=&   \left(  p_{\parallel} - p_o\right)^2 + 2 p_\perp^2  + U^2\bar m^2 \nonumber\\
&\pm&
2p_\perp\sqrt{
 \left(  p_{\parallel} - p_o\right)^2 + 2U^2\bar m^2
}
\end{eqnarray}
with two nodal lines at $p_{\parallel} - p_o=0$ and $\sqrt{2}p_\perp=\pm U\bar m$. 
Unlike the nodal lines in the first two cases, these two are tilted along $p_{\parallel}$ and move away from each other along $p_{\perp}$ for $\delta\neq 0$.
\end{enumerate}

\section{Example models}
\label{examples}

In this section we investigate the density wave phases in several specific lattice models
which have been widely studied in the literature. 
The examples in Secs. \ref{model1_example} and \ref{model2_example}  fall into the two types of NLs discussed in previous sections. 
The remainder of this section will be devoted to model beyond the simplest two-band cases.

\subsection{A model-1 loop example}\label{model1_example}

A lattice model with two ``model-1" loops can be described by the Hamiltonian,
\begin{eqnarray}
H_0(\bfk)&=& \left( \cos k_x + \cos k_y - m \right)\tau_a + \cos{k_z} \tau_b\label{model1}\,.
\end{eqnarray}
When $|m|<2$, the system has two NLs at $\cos{k_x}+\cos{k_y}-m=0$,  in two parallel planes $k_z=\pm \pi/2$.
 They are nested by the vector $\mathbf{Q}=(0,0,\pi)$, and can be mapped to each other as
\begin{eqnarray}
H_0(\bfk)=\tau_aH_0(\bfk+{\bf Q})\tau_a\,.
\end{eqnarray}
Introducing a Hubbard interaction, the effective Hamiltonian can be written as
\begin{eqnarray}
H_{eff}(\vk) &=&
 \left( \cos k_x + \cos k_y - m \right)t_0\tau_a +  \cos{k_z} t_3\tau_b \nonumber\\ 
 &+& U\bar m\  t_1\tau_{\alpha}\,.
\end{eqnarray}

We analyze first the CDW case ($\tau_\alpha=\tau_0$). 
Following the discussion in the previous Section, 
this condition corresponds to the second case  in Sec. \ref{model1_general}, where each NL gets split into two.
Indeed, from the commuting relations between different terms, we can obtain the energy dispersion as
\begin{equation}
E_{CDW}= \pm \sqrt{\cos^2 {k_z} + \left[
\cos k_x + \cos k_y - m \pm U\bar m
\right]^2
}\,,
\end{equation}
and the split NLs are given by
\begin{eqnarray}
\cos{k_z}  = 0\,,\mbox{and}\
\cos k_x + \cos k_y = m \pm U\bar m.
\end{eqnarray}

 In the PSDW case ($\tau_\alpha=\tau_3$), different choices of $a$ and $b$ in Eq. (\ref{model1}) will lead to different phases of the system. In such case, the effective Hamiltonian reads
\begin{eqnarray}
H_{eff}(\vk) &=&
 \left( \cos k_x + \cos k_y - m \right)t_0\tau_a +  \cos{k_z} t_3\tau_b \nonumber\\ 
 &+& U\bar m\  t_1\tau_3,
\end{eqnarray}
and the possibilities are summarized in Table \ref{tab1}, and listed explicitly as follows.

\begin{table}
\begin{center}
\begin{tabular}{c|c|c|c}
a$\backslash$ b & 1 & 2 & 3 \\
\hline
1 & X & loops & gapped \\
\hline
2 & loops & X & gapped\\
\hline
3 & metal & metal & X
\end{tabular}
\end{center}
\caption{Possibilites for $\tau_a$, $\tau_b$ matrices in the loop model (\ref{model1}). And the resulting different outcomes
of the PSDW phase.}
\label{tab1}
\end{table}

In the case of $\tau_a=\tau_3$, the energy spectrum reads
\begin{eqnarray}E_{PSDW}&=&\pm\sqrt{ 
\left( \cos k_x + \cos k_y - m \right)^2 +  \cos {k_z}  ^2 }\pm U\bar m 
 \nonumber\\
\end{eqnarray}
with uncorrelated $\pm$ signs. This is the first case in Sec. \ref{model1_general}, and the ordered phase is metallic.

If  $a,b\neq 3$,   the spectrum reads
\begin{eqnarray}
E_{PSDW}^2&=&  
\left( \cos k_x + \cos k_y - m \right)^2 + \left(    \cos{k_z}    \pm U\bar m  \right)^2\,,
\end{eqnarray}
where each original NL splits into two with different $k_z$, as in the third
 cases in Sec. \ref{model1_general}.

Finally, when $\tau_b=\tau_3$, the spectrum takes the form,
\begin{eqnarray} 
E^2&=& 
\left( \cos k_x + \cos k_y - m \right)^2 +  \cos {k_z} ^2 + U^2\bar m^2\,.
\end{eqnarray}
Thus the system is fully gapped by a nonzero $U \bar m$ and becomes an insulator, as in the fourth case in Sec. \ref{model1_general}.

\subsection{A model-2 loop example}\label{model2_example}

By including $\cos{k_z}$ in the $\tau_a$ term in Eq. (\ref{model1}), we obtain a system with ``model-2" loops, described by
\begin{eqnarray}
H_0(\bfk)= \left( \cos k_x + \cos k_y + \cos k_z- m \right)\tau_a + \cos k_z  \tau_b.\label{model2}
\end{eqnarray}
This system has two parallel NLs with $k_z=\pm\pi/2$ when $-2<m<2$.
The continuous approximation for these two loops, as in Eqs. (\ref{loop2}), satisfies $g'_1=1$, and $g_2=g'_2=-1$.
Thus, the effective Hamiltonian including the Hubbard interaction is given by
\begin{eqnarray}
H_{eff} &=&  \left( \cos k_x + \cos k_y - m \right)t_0\tau_a +   \cos k_z  t_3\tau_a\nonumber\\ &+&
\cos k_z  t_3\tau_b + U\bar m t_1\tau_{\alpha}\,.
\label{16}
\end{eqnarray}

 For the CDW case  ($\tau_{\alpha}=\tau_0$), the spectrum reads
\begin{eqnarray}
E^2_{CDW}&=& \left[
\cos k_x + \cos k_y  - m \pm \sqrt{U^2\bar m^2+\cos^2{k_z}}
\right]^2\nonumber\\
&&+\cos {k_z}^2 
\,,
\end{eqnarray}
thus each of the original NLs splits into two. The new NLs are given by
\begin{eqnarray}
\cos{k_z}  = 0\,, \hspace{0.2cm}
\cos k_x + \cos k_y = m \pm U\bar m.
\end{eqnarray}

In the case of   PSDW ordering ($\tau_{\alpha}=\tau_3$), 
there are three possible outcomes as summarized in Table \ref{tab2}.
In the case of $\tau_a=\tau_3$, the energy spectrum reads
$$
E^2_{PSDW}=(\cos k_x+\cos k_y-m)^2+2\cos^2 k_z+U^2\bar m^2
$$
\begin{eqnarray}
\pm2\sqrt{(\cos k_x+\cos k_y-m)^2\left[  \cos^2 k_z + U^2\bar m^2  \right] + (U\bar m\cos k_z)^2 }\,,\nonumber\\
\end{eqnarray}
where each NL splits into two as in the second case in Sec. \ref{model2_general}. 
The condition of the NLs is the same as for the CDW case;
 however, as discussed in the
 previous section [Fig. \ref{fig1}(b)], these NLs have a quadratic dispersion along $k_z$.

\begin{table}
\begin{center}
\begin{tabular}{c|c|c|c}
a$\backslash$ b & 1 & 2 & 3 \\
\hline
1 & X & loops & gapped \\
\hline
2 & loops & X & gapped\\
\hline
3 & loops & loops & X
\end{tabular}
\end{center}
\caption{Possibilites for $\tau_a$, $\tau_b$ matrices in the loop model (\ref{model2}). And the resulting different outcomes
of the PSDW phase.}
\label{tab2}
\end{table}

If  $\tau_b=\tau_3$ then the spectrum reads
\begin{eqnarray}
E_{PSDW}^2&=& \left[
\cos k_z\pm \sqrt{ U^2\bar m^2+ \left(   \cos k_x + \cos k_y -m\right)^2}
\right]^2\nonumber\\
 &+& \cos^2k_z \,,
\end{eqnarray}
which is an insulating phase as in the third case in Sec. \ref{model2_general}.

The remaining possibility, $a,b\neq 3$, where the spectrum reads 
\begin{eqnarray}E^2_{PSDW}&=&(\cos k_x+\cos k_y-m)^2+2\cos^2 k_z+U^2\bar m^2\nonumber\\
&&\pm2\cos k_z\sqrt{
(\cos k_x+\cos k_y-m)^2
+2U^2\bar m^2
}\,,\nonumber\\
\end{eqnarray}
yields two NLs  given by
\begin{eqnarray}
\cos{k_z}  = \pm U\bar m /\sqrt{2}\,, \
\cos k_x + \cos k_y = m\,.
\end{eqnarray}
This is the splitting along $k_z$, as in the fourth case in Sec. \ref{model2_general}. 
Finally, if we add an extra term $H_{\delta}=\delta\sin{k_z}\tau_0$ to the original Hamiltonian of Eq. (\ref{model2}), we can induce a energy offset $\delta$ of the two original NLs, and tilt the resulting NLs after the Hubbard interaction is introduced.

\subsection{Nested   $\mathbb{Z}_2$ NLs }
\label{sec:Z2}
We have hitherto considered examples of two-band NLSMs, which verify our results for general two-band models. If an extra degree of freedom, say, a (pseudo)spin-1/2 subspace, is introduced, the Hilbert space is enlarged and the possibility of nodal lines
carrying a $\mathbb{Z}_2$ monopole charge arises, which must be created in pairs\cite{Z2_loop}. 
It is beyond the scope of this paper to extend the  general analysis of  Sec III. 
Instead, we explicitly consider  a recent  four-band  model for $Z_2$ NLs \cite{Z2_loop}  
and  study  density-wave order due to nesting. 
The model reads
\begin{eqnarray}
H_0(\bfk) &=&  \sin k_x \tau_0\sigma_1+ \sin k_y \tau_2\sigma_2+ 
\sin k_z   \tau_0\sigma_3 + m \tau_1\sigma_1\,,\nonumber\\ 
\label{Z2_H}
\end{eqnarray}
and has the spectrum
\begin{eqnarray}
E_0^2(\vk)&=&
\left(   \sqrt{\sin^2k_x + \sin^2k_y}\pm m  \right)^2 + \sin^2k_z\,.\label{spectrum_Z2}
\end{eqnarray}
There are  eight NLs,  centered at momenta $\vk$ with
Cartesian components
 $k_i=0,\pi$   for $i=x,y,z$. 
We introduce a Hubbard interaction in  four-dimensional space assuming that the repulsion exists
between the two orbitals in subspace of $\boldsymbol\sigma$:
\begin{eqnarray}
\hat U&=&  U  \sum_{\br , \nu}n_{\nu 1}(\br) n_{\nu 2}(\br) = \frac{U}{2}  \sum_{\br}  n(\br)\tau_0\sigma_1 n(\br)\,,
\end{eqnarray}
so that the remaining index  $\nu=1,2$ for the two components in the
subspace of  $\boldsymbol\tau$
 produces a two-fold degeneracy.
In the mean-field approximation, the interaction reads (apart from unimportant constants)
\begin{eqnarray}
\hat U_{eff}&=&  U\bar m \sum_{\vk} \hat c_{\vk+\vQ}^\dagger \tau_0\sigma_z \hat c_{\vk} \,,
\end{eqnarray}
where we defined the PSDW as:
 \begin{equation}
\langle   n_{\nu j}(\br)\rangle = \frac{n}{2} + \bar m (-1)^j\cos(\vQ\cdot\br) \,,
\end{equation}
The effective 8x8  Hamiltonian has the same form as that in  Eq. (\ref{martizHloops}), but the anti-diagonal  blocks 
are now written as  $U\bar m \tau_0\sigma_3$. 
Two of the original $\mathbb{Z}_2$ loops
are now coupled by the interaction. 
Such coupling provides an extra pseudospin-1/2 subspace, and the interaction term breaks the SU(2) symmetry in this space.
As a result, the pair of NLs may either survive, shrink to point nodes, or  be  gapped out by the PSDW.

Depending on vector $\vQ$, the effective Hamiltonian takes the form of
\begin{eqnarray}
H_{eff} &=& \sin k_x t_{\alpha_x} \tau_0\sigma_1 + \sin k_y t_{\alpha_y} \tau_2\sigma_2 + \sin k_z t_{\alpha_z} \tau_0\sigma_3 \nonumber\\ 
&+&m t_0 \tau_1\sigma_1 + U\bar m t_1 \tau_0\sigma_3\,,
\end{eqnarray}
where the Pauli matrix $t_{\alpha_i}$ equals $t_0$ ($t_3$) for $Q_i=0$ ($Q_i=\pi$). 
Thus, different choices of $Q_i$ determine whether each term commutes or anticommutes with the interaction term 
$U\bar m t_1 \tau_0\sigma_3$.
The possibilities with different choices of $\vQ$ are summarized as follows.
\begin{figure}
\includegraphics[width=1\linewidth]{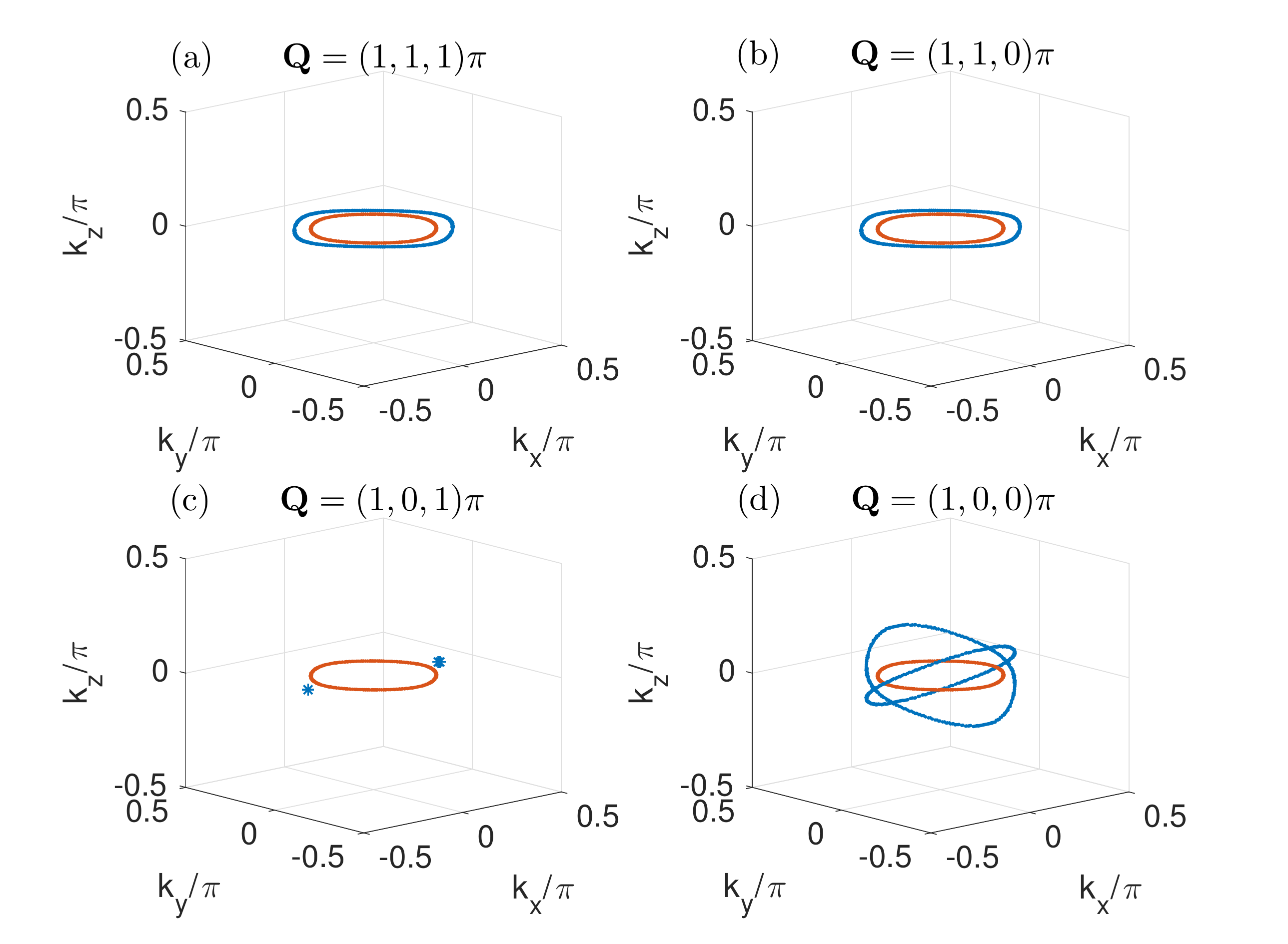}
\caption{The effect of a PSDW on the $\mathbb{Z}_2$ loops for different nesting vectors $\vQ$. (a) to (d) show four different cases of $\vQ$ where the resulting system is still a semimetal. The red lines are the original NLs of Hamiltonian $H_0(\bfk)$ in Eq. (\ref{Z2_H}), whereas the blue lines and stars are the nodal lines or points  in  the PSDW phase.
The parameters are:  $m=\sqrt{2}/2$ and $U\bar m=0.5$. 
We only plot for  $k_i\in [-\pi/2,\pi/2]$, as the graph repeats itself with a period of $\pi$.\label{fig2}}
\end{figure}
\begin{enumerate}
\item $\vQ=(1,1,1)\pi$:
\begin{eqnarray}
H_{eff} &=& \sin k_x t_3 \tau_0\sigma_1 + \sin k_y t_3 \tau_2\sigma_2 + \sin k_z t_3 \tau_0\sigma_3 \nonumber\\ &+&
m t_0 \tau_1\sigma_1 + U\bar m t_1 \tau_0\sigma_3\,, 
\\
E^2 &=& \left(  \sqrt{\sin^2 k_x + \sin^2k_y} \pm  \sqrt{m^2  +U^2\bar m^2}\right)^2 + \sin^2k_z\,. \nonumber\\ 
\label{E111}
\end{eqnarray}
The spectrum is composed of four two-fold degenerate bands, and 
has the same NLs as the original $H_0$, albeit with enlarged radius [Fig. \ref{fig2}(a)]:
\begin{eqnarray}
k_z=0,\pi; \   \sqrt{\sin^2 k_x + \sin^2k_y} =  \sqrt{m^2  +U^2\bar m^2}\,.
\label{eq56}
\end{eqnarray}
Since the  energy dispersions in Eq.  (\ref{E111}) are two-fold degenerate,  the NLs are four-fold degenerate.

\item $\vQ=(1,1,0)\pi$:
\begin{eqnarray}
H_{eff} &=& \sin k_x t_3 \tau_0\sigma_1 + \sin k_y t_3 \tau_2\sigma_2 + \sin k_z t_0 \tau_0\sigma_3  \nonumber\\ &+&
m t_0 \tau_1\sigma_1 + U\bar m t_1 \tau_0\sigma_3\,,
\end{eqnarray}
which has the same NLs as in Eq. (\ref{eq56}), and
as shown in FIg. \ref{fig2}(b), despite some minor variation of the spectrum near the NLs.

\item $\vQ=(1,0,1)\pi$:
\begin{eqnarray}
H_{eff} &=& \sin k_x t_3 \tau_0\sigma_1 + \sin k_y t_0 \tau_2\sigma_2 + \sin k_z t_3 \tau_0\sigma_3 \nonumber\\ &+&
m t_0 \tau_1\sigma_1 + U\bar m t_1 \tau_0\sigma_3\,.
\end{eqnarray}
The full spectrum also has the  two-fold degeneracy. while the NLs are gapped in most regions, leaving only pairs of four-fold degenerate points at
\begin{eqnarray}
\sin k_y = \sin k_z=0; \   \sin k_x  = \pm \sqrt{m^2  +U^2\bar m^2}\,,
\end{eqnarray}
as shown in Fig. \ref{fig2}(c).

We note that the case $\vQ=(0,1,1)\pi$ has a similar spectrum, which can be obtained from 
the above  by performing the substitution $k_x\leftrightarrow k_y$.

\item 
$\vQ=(1,0,0)\pi$:
\begin{eqnarray}
H_{eff} &=& \sin k_x t_3 \tau_0\sigma_1 + \sin k_y t_0 \tau_2\sigma_2 + \sin k_z t_0 \tau_0\sigma_3 \nonumber\\ &+&
m t_0 \tau_1\sigma_1 + U\bar m t_1 \tau_0\sigma_3\,.
\end{eqnarray}
Energy zeros are obtained if  two conditions are simultaneously satisfied:
\begin{subequations}
\begin{eqnarray}
\sum_{i=1}^3 \sin^2k_i =  U^2\bar m^2 + m^2\\
m\sin k_z=\pm  U\bar m \sin k_y\,.
\end{eqnarray}
\end{subequations}
Geometrically, one can think of the first condition as defining 
a spherical surface, for small $\vk$, and the second one as two planes.
The intersection between the surface and the planes yields two NLs, 
obtained by rotating the original loops (in the XY plane) around the $x$ axis in opposite directions.  
These two NLs are given by  different pairs of energy bands, and  thus form a nodal chain [Fig. \ref{fig2}(d)].

We note that spectrum for  the case  $\vQ=(0,1,0)\pi$
 is similar to the  case $\vQ=(\pi,0,0)$, and differs only in the interchange $k_x\leftrightarrow k_y$.
The nodal chain is obtained from the  two original NLs  by a rotation around the $y$ axis.  

\item $\vQ=(0,0,1)\pi$
\begin{eqnarray}
H_{eff} &=& \sin k_x t_0 \tau_0\sigma_1 + \sin k_y t_0 \tau_2\sigma_2 + \sin k_z t_3 \tau_0\sigma_3 \nonumber\\ &+&
m t_0 \tau_1\sigma_1 + U\bar m t_1 \tau_0\sigma_3\,.
\end{eqnarray}
In this case the system is  a fully gapped  insulator.
\end{enumerate}

 \section{Nested NLs in Dirac systems}
 \label{Diracsec}

In this Section we study density wave phases in four-band spinfull  Hamiltonians
 with NLs.  The spin degree of freedom allows us to distinguish two types of ordered phases:
 true and hidden spin density waves (SDWs).  In subsection \ref{Diracperturb} we  consider NLs
 obtained from a perturbed\cite{Burkov,mitschell}  Dirac Hamiltonian.
  
 \subsection{Spin degenerate loops}

The simplest way to go from a Weyl to a Dirac loop is to introduce  spin degeneracy,
 \begin{equation}
H_0(\vk)\  \rightarrow\ H_0(\vk) \sigma_0
\end{equation}
where $\sigma_\mu$ acts in spin space.
 
 We assume Hubbard repulsion between two fermions having opposite spins in the same orbital
(labeled by the index $j$):
\begin{eqnarray}
\hat U = \sum_{\br, j=1,2} \hat n_{j\up}(\br)\hat n_{j\dn}(\br)\,.
\label{Usigma}
\end{eqnarray}
 In principle one could have  ordered phases
 with ferromagnetic (FM) or antiferromagnetic (or SDW) configurations of the spin:  
\begin{eqnarray}
\langle n_{j\sigma}\rangle &=& 
  \frac 1 4 n + \bar m  \sigma  \qquad
  \sigma=\pm 1\,,\qquad\mbox{Stoner FM}\\
  \langle n_{j\sigma}\rangle &=& 
  \frac 1 4 n + \bar m \sigma (-1)^j   \qquad\mbox{hidden Stoner FM}\\
 \langle n_{j\sigma}\rangle &=& 
  \frac 1 4 n + \bar m  \sigma \cos(\vQ\cdot\br) \qquad\mbox{true SDW}\\
  \langle n_{j\sigma}\rangle &=& 
  \frac 1 4 n + \bar m  \sigma (-1)^j \cos(\vQ\cdot\br)   \qquad\mbox{hidden  SDW} \label{hiddenSDW}
 \end{eqnarray}
 Because a NL's density of states vanishes linearly with energy, 
 the Stoner criterion precludes the FM orderings for weak interactions\cite{roy}, 
and they are not related  to the nesting $\vQ$.
In the following, we shall concentrate on SDW  phases.
Considering the hidden SDW, Eq. (\ref{hiddenSDW}),   the effective interaction reads: 
\begin{eqnarray}
\hat U_{eff} &=&
-U\sum_{\br,j} \left[ \left( \frac n 4  \right)^2  -   \bar m^2  \cos^2(\vQ\cdot\br) 
- \frac n 4 \sum_{\sigma} \hat\psi^\dagger_{j\sigma}(\br) \hat\psi_{j\sigma}(\br) 
\right]\nonumber\\
&+&U\bar m\sum_{\br,j,j',\sigma\sigma'} \hat\psi^\dagger_{j\sigma}(\br)\
  \tau^{jj'}_3\sigma^{\sigma\sigma'}_3  \cos(\vQ\cdot\br)\   \hat\psi_{j'\sigma'}(\br) \,,\label{tSDW}
\end{eqnarray}
where the field operator $\hat\psi_{j\sigma}(\br)$ now includes the spin index $\sigma$. 
Similarly, the
 field operator in momentum space $\boldsymbol c_\bfk$ now denotes $ c_{\bfk,j,\sigma}$ for all $j,\sigma$.
The Hamiltonian matrix  in $\left(  \boldsymbol c^\dagger_{\bfk}\ \boldsymbol c^\dagger_{\bfk+\vQ}  \right)$ space reads
(apart from unimportant constants),
 \begin{eqnarray}
H_{eff}(\vk) &=&  \left(  \begin{array}{cc}
H_0(\vk) & U\bar m \tau_\alpha\sigma_3 \\
 U\bar m \tau_\alpha\sigma_3  & H_0(\vk+\vQ)  
\end{array}\right)\,,\label{SDWDirac}
\end{eqnarray}
where $\alpha=3$ describes a hidden SDW [as in Eq. (\ref{tSDW})],
and   $\alpha=0$ describes a true SDW.
The off-diagonal block $U\bar m t_1\tau_3\sigma_3$ has exactly the same (anti)commutation 
relations with the other
Hamiltonian terms, 
as in the Weyl case of Sections \ref{model1_general} and \ref{model2_general} . 
All the criteria and spectra established for the Weyl case still
hold, if one just replaces  $\bar m\rightarrow\bar m \sigma_3$. 
Since this term appears as $(U\bar m)^2$ in the dispersion relations,  there is no spin splitting in the spectra.
 
We note that single antiferromagnetic NLs have been discussed in the literature\cite{jingwang}.
 
 \subsection{Two perturbed Dirac points}
 \label{Diracperturb}

 A nodal line Dirac semimetal can be obtained starting from a pristine 3D Dirac semimetal\cite{mitschell}
 of the form
 $H_D(\vk)  = -\tau_3\vp\cdot\boldsymbol\sigma$ and perturbing it with terms of the form
 $a_\mu\tau_\mu\otimes b_\nu\sigma_\nu$.  
 Suppose, for instance, 
 \begin{equation}
H_0(\vk)  = -\tau_3\vp\cdot\boldsymbol\sigma + \tau_1\boldsymbol b\cdot\boldsymbol\sigma
\end{equation}
 Without loss of generality assume $\boldsymbol b \parallel \hat z$. The term
$p_z\tau_3\sigma_3$ anticommutes with the others, so:
 \begin{equation}
 E^2 =p_z^2 + \left( \sqrt{p_x^2+p_y^2}\pm b   
 \right)^2 \,.
 \end{equation}
We note that this dispersion relation is very similar to that in Eq. (\ref{spectrum_Z2}) for the $\mathbb{Z}_2$ loops. 
However, the effect of the Hubbard interaction is different, as these two systems couple the two sets
 of (pseudo)spin-1/2 subspace in different ways. On the other hand,
Dirac points described by $H_D$ above
 do not exist alone if an additional symmetry, such as time-reversal or inversion, is present.
 For instance, a time-reversal symmetry (TRS) relates two  Dirac points at $-\frac 1 2 \vQ$ and  $+\frac 1 2 \vQ$ 
  in such a way that:
 \begin{eqnarray}
\sigma_2H_0^*\left(\frac \vQ 2 -\bfk\right)\sigma_2 = H_0\left(-\frac \vQ 2 +\bfk\right)\,,
\end{eqnarray}
 it then follows that,  for $\boldsymbol b=0$,
$H_0\left(- \vQ /2 +\bfk\right) =H_0\left( \vQ /2 +\bfk\right) =H_D(\bfk)$.
Therefore, the two unperturbed Dirac points have the same $k\cdot p$ Hamiltonian.
Including  the $\tau_1\boldsymbol b\cdot\boldsymbol\sigma$ term, which breaks TRS, 
we  obtain the model,
 \begin{subequations}
 \begin{eqnarray}
 H_0\left(-\frac \vQ 2 +\bfk\right) &=& -\tau_3\vp\cdot\boldsymbol\sigma +\tau_1\boldsymbol b\cdot\boldsymbol\sigma\,,\\
 H_0\left(\frac \vQ 2 +\bfk\right) &=& -\tau_3\vp\cdot\boldsymbol\sigma +\tau_1\boldsymbol b\cdot\boldsymbol\sigma\,,
\end{eqnarray}\label{Tpartial}
  \end{subequations}
 and the effective Hamiltonian has equal diagonal blocks.

A different version of the above model that would preserve TRS symmetry reads: 
 \begin{subequations}
 \begin{eqnarray}
 H_0\left(-\frac \vQ 2 +\bfk\right) &=& -\tau_3\vp\cdot\boldsymbol\sigma +\tau_1\boldsymbol b\cdot\boldsymbol\sigma\,,\\
 H_0\left(\frac \vQ 2 +\bfk\right) &=& -\tau_3\vp\cdot\boldsymbol\sigma -\tau_1\boldsymbol b\cdot\boldsymbol\sigma\,.
\end{eqnarray}\label{Tcase}
  \end{subequations}
  
 If we now consider the role of inversion symmetry $\bfk\rightarrow -\bfk$, 
 the two Dirac points are related by 
  \begin{subequations}
 \begin{eqnarray}
H_0\left(\frac \vQ 2 -\bfk\right) = H_0\left(-\frac \vQ 2 +\bfk\right) &=& -\tau_3\vp\cdot\boldsymbol\sigma + \tau_1\boldsymbol b\cdot\boldsymbol\sigma\,,\hspace{1cm} \\ 
\Rightarrow
H_0\left(\frac \vQ 2 +\bfk\right) &=& \tau_3\vp\cdot\boldsymbol\sigma +\tau_1\boldsymbol b\cdot\boldsymbol\sigma\,,
\label{Pcase}
\end{eqnarray}
  \end{subequations}
therefore, the two Dirac points have  different $k\cdot p$ Hamiltonians.

Next we study the effects of a hidden SDW and a true SDW for these different cases, 
still assuming $\boldsymbol b \parallel \hat z$. 
For a hidden SDW,
   the effective Hamiltonian for the TRS  breaking  model in Eq. (\ref{Tpartial})  is then
 \begin{eqnarray}
\hat H_{eff} =  t_0 \left[-\tau_3\vp\cdot\boldsymbol\sigma + b\tau_1 \sigma_3\right] +  U\bar m t_1\tau_3\sigma_3\,,
\end{eqnarray}
which, by inspection produces the eight band spectrum:
\begin{equation}
 E^2 = \left( - p_z\pm U\bar m \right)^2 + \left(  \sqrt{p_x^2+p_y^2}\pm b   \right)^2
 \end{equation}
 where the $\pm$ signs are uncorrelated.  This corresponds to splitting each  loop along the $k_z$ direction. 
 
If one considers,  instead, a  true SDW phase,  
  \begin{eqnarray}
\hat H_{eff} =  t_0 \left[-\tau_3\vp\cdot\boldsymbol\sigma + b\tau_1\sigma_3\right]+  U\bar m t_1\tau_0\sigma_3\,,
\end{eqnarray}
then there are four doubly degenerate bands:
 \begin{eqnarray}
E^2&=& b^2 +  U^2\bar m^2+ \vp^2\nonumber\\ &\pm& 2\sqrt{ b^2\left(  U^2\bar m^2 + p_x^2 + p_y^2  \right) + U^2\bar m^2p_z^2
} \label{mathspec}
\end{eqnarray}
with nodal lines given by   $ p_z=0\,,  p_x^2 + p_y^2  = b^2 - U^2\bar m^2$.
So, the initial two loops still exist
 but their radius shrinks.

  In the TRS model,  Eq. (\ref{Tcase}), the hidden SDW phase is described by the effective Hamiltonian:
  \begin{eqnarray}
\hat H_{eff}&=&  -t_0\tau_3\vp\cdot\boldsymbol\sigma + bt_3\tau_1\sigma_3 +  U\bar m t_1\tau_3\sigma_3\,.
\end{eqnarray}
The spectrum is  the same as in Eq. (\ref{mathspec}).
 So, the initial two loops still exist but their radius shrinks.
 And a true SDW phase is described by the effective Hamiltonian:
 \begin{eqnarray}
\hat H_{eff} &=&  -t_0 \tau_3( p_x\sigma_1 + p_y\sigma_2) - p_z t_0\tau_3\sigma_3 + b\   t_3\tau_1\sigma_3 \nonumber\\ &+& U\bar m t_1\tau_0\sigma_3\,,
\end{eqnarray}
 which produces the  spectrum with eight bands:
 \begin{equation}
 E^2 =\left(    \sqrt{p_x^2+p_y^2} \pm b  \right)^2 + \left( p_z \pm     U\bar m  \right)^2\,,
 \end{equation}
 where the $\pm$ signs are uncorrelated.  This corresponds to splitting each  loop along $p_z=\pm U\bar m$. 
 
 For the case with inversion symmetry, in Eq. (\ref{Pcase}), a hidden SDW phase is described by the effective Hamiltonian:
 \begin{eqnarray}
H_{eff} (\bfk)=  -t_3 \tau_3\vp\cdot\boldsymbol\sigma + b \ t_0\tau_1\sigma_3 +  U\bar m t_1\tau_3\sigma_3\,,
\label{PhiddenSDW}
\end{eqnarray}
which produces the eight band spectrum:
 \begin{equation}
 E^2 =p_z^2 +  \left(    \pm \sqrt{b^2 + U^2\bar m^2} \pm   \sqrt{p_x^2+p_y^2}
  \right)^2\,,
 \end{equation}
  (uncorrelated$\pm$ signs).  This corresponds to splitting each nodal  line by changing its radius.
 A true SDW is obtained by changing $\tau_3\rightarrow\tau_0$ in the last term of Eq. (\ref{PhiddenSDW}).
The resulting spectrum, \
  \begin{equation}
 E^2 =p_z^2 +  \left(   \sqrt{p_x^2+p_y^2} \pm b \pm  U\bar m   \right)^2\,,
 \end{equation}
 (with uncorrelated $\pm$ signs) also has NLs given by $p_z=0$, $\sqrt{p_x^2+p_y^2} =|b\pm  U\bar m|$.
 
In the remaining case, where  $H_0(-\vQ/2 + \bfk)=-H_0(\vQ/2  + \bfk)$,
a  hidden SDW phase  is described by the effective Hamiltonian:
 \begin{eqnarray}
H_{eff} (\bfk)=  t_3 \left[ -\tau_3\vp\cdot\boldsymbol\sigma + b \tau_1\sigma_3 \right]+  U\bar m t_1\tau_3\sigma_3\,,
\end{eqnarray}
which, by inspection produces the eight band spectrum: 
 \begin{equation}
 E^2 =p_z^2 +  \left(    \sqrt{p_x^2+p_y^2} \pm b \pm  U\bar m \right)^2\,,
 \end{equation}
 with uncorrelated$\pm$ signs.  Therefore,  each  loop splits in the radial direction. The
  true SDW is described by the effective Hamiltonian:
 \begin{eqnarray}
\hat H_{eff} = t_3 \left[ -\tau_3\vp\cdot\boldsymbol\sigma + b \tau_1\sigma_3 \right]+  U\bar m t_1\tau_0\sigma_3\,,
\end{eqnarray}
 which, by inspection produces the eight band spectrum:
 \begin{equation}
 E^2 = p_z^2 +  \left(  \pm  \sqrt{p_x^2+p_y^2} \pm \sqrt{ b^2 + U^2\bar m^2}  \right)^2 \,,
 \end{equation}
 where the $\pm$ signs are uncorrelated.  Again, this corresponds to splitting each nodal  loop in the radial direction.

\section{Superconductivity}
 \label{supersec} 
 
 When considering a single Weyl NL, the pairing block of the Bogolyubov-deGennes
 (BdG) matrix
 in the   particle-hole   basis\cite{sacramento1,sacramento2,beri}, takes the form
 \begin{equation}
\hat \Delta(\vk) = 
 \left[ 
d_0(\vk)\tau_0+ \boldsymbol d(\vk)\cdot\boldsymbol\tau
 \right] i\tau_2
\,,
\end{equation}
and fermionic  statistics imposes that  
$\hat \Delta(\vk) = \hat \Delta^T(-\vk)$.
  Close to the nodal lines the   3D momentum, $\vp=\hbar\vk$,
 can be parametrized as 
 \begin{subequations}
\begin{eqnarray}
  p_x &=& (p_0 + \tilde p\cos\phi) \cos\theta\\ 
  p_y &=& (p_0 + \tilde p\cos\phi) \sin\theta \,,\\
p_z&=&p_\perp=\tilde p\sin\phi\,, 
\end{eqnarray}\label{pcoordinates}
\end{subequations}
which is to be inserted in the $k\cdot p$ loops models. 
Here, $\theta$ is the azimuthal angle along the loop, 
$\tilde p$ is the radius of a torus involving  the NL, and the angle $\phi$ wraps around
the latter\cite{Nandkishore1}.
Note that, according to Eq. (\ref{pcoordinates}),  momentum inversion $\vp\rightarrow -\vp$ is equivalent to
$\theta\rightarrow\theta+\pi$ and $\phi\rightarrow -\phi$, while reflection in the loop's
plane, $p_z\rightarrow -p_z$, is equivalent to $\phi\rightarrow -\phi$.

In the semimetal case (undoped, or compensated case) the FS reduces to the  NL
and $\vp$ reduces to the angle $\theta$ on the loop. In the doped case, any point on 
the  torus shaped FS can be labeled by two angles, $\theta,\phi$. The functions
$ d_0(\vk)$  and $\boldsymbol d(\vk)$ describe (pseudo-spin) singlet  and 
triplet pairing, respectively. 
 One can expand the singlet pairing function  quite generally as
\begin{equation}
d_0(\vk)  =  \sum_{l_1,l_2} e^{il_1\theta}\left[  \Delta_{l_1l_2}\cos(l_2\phi) + \tilde \Delta_{l_1l_2}\sin(l_2\phi)
\right]\,.
\end{equation}
An analogous expansion can be written for $\boldsymbol d(\vk)$.

If there are two nested Weyl loops, then an additional loop label
must be introduced and the Pauli matrix $t_\mu$ operates in the two-dimensional loop space. For a two-loop
system then, we write the  pairing matrix as
\begin{equation}
\hat \Delta(\vk) = 
 \left[ 
d_0(\vk)\tau_0+ \boldsymbol d(\vk)\cdot\boldsymbol\tau
 \right] i\tau_2
 t_\mu\,.
\end{equation}
 The BdG Hamiltonian matrix in the particle-hole basis has the form
 \begin{equation}
H(\vk)= \left( \begin{array}{cc}
\hat\Xi (\vk)& \hat \Delta (\vk)\\
\hat\Delta^\dagger (\vk) & -\hat\Xi^T(-\vk)
 \end{array}\right) 
 \label{BdGmatrix}
\end{equation}
with $\hat\Xi= diag(H_1, H_2)$ .  The Hamiltonians $H_{1(2)}$ are the $k\cdot p$ Weyl NL models.
The total Hamiltonian is then 
\begin{eqnarray}
\hat H=\frac 1 2  \sum_{\vk}{\boldsymbol c}^\dagger H(\vk)
 {\boldsymbol c}\,,
\end{eqnarray}
where $ {\boldsymbol c}=(\hat {\boldsymbol c}_{\vk,1},\   \hat  {\boldsymbol c}_{\vk,2},\ 
 \hat  {\boldsymbol c}_{-\vk,1}^\dagger,\   {\boldsymbol c}_{-\vk,2}^\dagger
)^T$.
If the two NLs are centered at BZ points $\pm\vQ/2$ respectively, then the inter-NL pairing
is the ``usual'' pairing between opposite momenta, and we shall take this  to be the case.
 If not, then the Cooper pair would have 
a finite quasi-momentum (a Fulde-Ferrel-Larkin-Ovchinnikov  state\cite{LOFF1,LOFF2,LOFF3}).

The cases  $\mu=0,1,3$, are different from the case $\mu=2$ regarding the parity of the functions 
$ d_0(\vk)$ and $\boldsymbol d(\vk)$.
In the cases $\mu=1,2$,   electrons
on different loops are being paired: an electron $(\vk ,1)$ is being paired with another $(-\vk ,2)$.
The cases $\mu=0,3$ describe intra-NL  pairing, where the scattering of two particles from one NL 
into the other may be included,  and $id_0\tau_2t_3$ would describe sign-reversed s-wave 
 pairing, analogous to that in pnictide superconductors\cite{BangChoi}. 
 
Inter-NL pairing with $\mu=1$ (interloop triplet pairing), requires
$d_0$ to be even and ${\boldsymbol d}$ to be odd  function of $\vk$;  if $\mu=2$ (interloop singlet), then 
$d_0$ and ${\boldsymbol d}$ have the opposite parities. 
The BdG matrix decouples into two blocks
each associated with
 the vector spaces   $(\hat c_{\vk,1}, \hat c_{-\vk,2}^\dagger)^T$
and $(\hat c_{\vk,2}, \hat c_{-\vk,1}^\dagger)^T$, respectively.
Since we expect a  fully gapped excitation spectrum to have higher condensation energy than a nodal spectrum, 
we shall examine the cases where $d_0$ and $\boldsymbol d$ are constant on the FS (in the
$\mu=1$ and $\mu=2$ cases, respectively). If TRS holds, then these order parameters must also be real.

\subsection{Model-1 loops}

Assuming a positive energy offset, $\delta$, the interband pairing occurs
between the electronic toroidal FS from the $H_1-\delta$ loop,
and the hole-like FS  from the $H_2+\delta$ loop.  
As in   previous literature, this is best done by considering projective form factors\cite{wangye,Nandkishorepro}
 onto the conduction or valence band. 
Let $U_{1(2)}$ be the unitary matrices which diagonalize $H_{1(2)}$, so that 
$U_sH_sU_s^\dagger=\sqrt{\left( |\vp_\parallel| - p_0 \right)^2 + p_\perp^2}\ \tau_3\equiv  \tilde p\tau_3$ for $s=1,2$. 
The positive and the
negative branches are the conduction and valence bands, respectively.
Because for  model-1 loops there is always a Pauli matrix $\tau_\beta$ such that 
 $H_1=\tau_\beta H_2\tau_\beta$, 
 it then follows that   $U_2=U_1\tau_\beta$.
We can apply this same unitary transformation to the BdG matrix in Eq. (\ref{BdGmatrix}) as:
\begin{eqnarray}
\left( \begin{array}{cc}\Lambda & 0\\ 0 & \Lambda^*(-\vk)\end{array}\right)H(\vk) 
\left( \begin{array}{cc}\Lambda^\dagger & 0\\ 0 &\Lambda^T(-\vk) \end{array}\right)
\nonumber\\ =
\left( \begin{array}{cc} \tilde p\tau_3t_0 -\delta\tau_0 t_3 & \Lambda \hat\Delta \Lambda^T(-\vk) \\ 
\Lambda^*(-\vk)  \hat\Delta^\dagger \Lambda^\dagger& -\tilde p\tau_3t_0+\delta \tau_0 t_3 \end{array}\right)
\label{UBdGU}
\end{eqnarray}
where $\Lambda=diag(U_1, U_2)$.
The off-diagonal pairing block is then $\Lambda(\vk) \hat\Delta(\vk) \Lambda^T(-\vk)$.

For $\delta>0$,  only the pairing between the conduction band of $H_1$ and the valence band of $H_2$ is considered. 
From the  BdG matrix in Eq. (\ref{UBdGU})  we obtain the submatrix operating in this two-fold subspace as:
\begin{eqnarray}
H^{FS}=\left( \begin{array}{cc} \tilde p -\delta & 
\Delta_{FS}(\vk) \\ 
\Delta_{FS}^*(\vk)
& \tilde p -\delta  \end{array}\right)
\label{condval1}
\end{eqnarray}
where $\Delta_{FS}(\vk)$ is
 the pairing  function on the FS which, from Eq. (\ref{UBdGU}) and  for  $\mu=1$  reads:
  \begin{eqnarray}
\Delta_{FS}(\vk) = 
 \left[
 U_1(\bfk)  \left(d_0+{\boldsymbol d}\cdot{\boldsymbol\tau}\right) i\tau_2U_2^T(-\bfk)\right]_{12}
 \label{pairingfunction}
\end{eqnarray}
It is then clear from Eq. (\ref{condval1}) that the spectrum  is $E=\tilde p -\delta \pm |\Delta_{FS}(\vk)|$, and is gapless.
At finite doping, no gapped state is to be expected from 
FS interloop pairing  between non-degenerate model-1 Weyl loops.

The situation is different for the degenerate ($\delta=0$) case, however, where the FS is composed of two nodal lines.
From Eq. (\ref{UBdGU}) and $t_\mu=t_1$ we obtain a  BdG matrix restricted to  the subspace $\left(  U_1(\vk)  \hat{\boldsymbol c}_{\vk,1} \,,    
U_2^*(-\vk) \hat{\boldsymbol c}_{-\vk, 2}^\dagger     \right)$, as
\begin{eqnarray}
H_{12}'=\left( \begin{array}{cc}
\tilde p\tau_3  &  U_1(d_0 + {\boldsymbol d}\cdot{\boldsymbol \tau})i\tau_2  U_2^T  \\
-iU_2^*\tau_2 (d_0 + {\boldsymbol d}\cdot{\boldsymbol \tau}) U_1^\dagger  & -\tilde p\tau_3 
\end{array}\right)
\label{UBU2}
\end{eqnarray}
For the sake of definiteness we consider the NL models with $\tau_a=\tau_1,\tau_b=\tau_2$, so that
  \begin{eqnarray}
  H_1(\phi) &=& \tilde p\left( \cos\phi\tau_1 + \sin\phi\tau_2\right)\,,\label{h1phi}\\
U_1(\vk)&=& \frac{1}{\sqrt{2}}\left( \begin{array}{cc}
1 & e^{-i\phi}\\
-1 & e^{-i\phi}
\end{array}\right)\,.\label{u1phi}\
\end{eqnarray}
We note that  all the other $(\tau_a,\tau_b)$ cases can be related to this through a suitable rotation in pseudo-spin  space.
From Eq. (\ref{u1phi}), one can see that $U_1^\dagger(\vk) =U_1^T(-\vk) $.
Interloop pairing is described by the off-diagonal block in  Eq (\ref{UBU2}):
  \begin{eqnarray}
 && U_1(\vk)  \left(d_0+{\boldsymbol d}\cdot{\boldsymbol\tau}\right) i\tau_2\tau_\beta^TU_1^T(-\vk)=\nonumber\\
  &=& 
 \left(\begin{array}{cc}
 id_0\sin\phi + id_2+d_3\cos\phi &  d_0\cos\phi + d_1 + id_3\sin\phi\\
  -d_0\cos\phi + d_1 - id_3\sin\phi &  - id_0\sin\phi + id_2-d_3\cos\phi 
\end{array}\right) \,, \nonumber \\
 &=&  \left(\begin{array}{cc}
d_3 +id_2\cos\phi- id_1\sin\phi  & -d_0 - d_1\cos\phi- d_2\sin\phi  \\ 
-d_0 + d_1\cos\phi + d_2\sin\phi  & d_3 -id_2\cos\phi+  id_1\sin\phi \end{array}\right) \,, \nonumber\\
 &=& \left(\begin{array}{cc} 
 -id_0 -id_1\cos\phi-id_2\sin\phi   &   d_1\sin\phi - d_2\cos\phi +id_3\\
 - d_1\sin\phi + d_2\cos\phi +id_3 &  -id_0 +id_1\cos\phi+id_2\sin\phi   \end{array}\right)\,,  \nonumber\\
 &=&
  \left(\begin{array}{cc} 
-d_0\cos\phi  - d_1-id_3\sin\phi &   -id_0\sin\phi - id_2 -d_3\cos\phi\\
  id_0\sin\phi - id_2 +d_3\cos\phi &  d_0\cos\phi  - d_1+id_3\sin\phi  \end{array}\right)\,,\nonumber\\
  \label{udelu}
\end{eqnarray}
for the cases $\beta=0,1,2,3$, respectively. Note that for the case $t_\mu=t_2$, 
we simply have to multiply
both sides of Eq. (\ref{udelu}) by $-i$.

For $\mu=1$ a fully gapped FS can only happen  for constant $d_0$ because $\boldsymbol d$ is an
odd function and must have nodes on the NLs.  In this case, only for $\beta=2$ a gapped
spectrum is obtained: $E^2 = \tilde p^2 + d_0^2$.

For $\mu=2$ (interloop singlet) and constant real $\boldsymbol d$ there are more possibilities.
If 
 $\beta=0$ a  fully gapped spectrum $E^2 = \tilde p^2 + d_2^2$; if
$\beta=1$ a fully gapped spectrum $E^2 = \tilde p^2 + d_3^2$; for   
$\beta=3$ the fully gapped spectrum $E^2 = \tilde p^2 + d_1^2$.
Gapped spectra result from  intraband  pairing. 
Interband pairing leads to nodal spectra for the same reason as  in
the $\delta>0$ case.

\subsection{Model-2 loops}

In a model-2 loop we replace Eqs. (\ref{h1phi})-(\ref{u1phi}) with
  \begin{eqnarray}
  H_1(\phi) &=& \tilde p\left[ (\cos\phi+ \sin\phi)\tau_1 + \sin\phi\tau_2\right]\,,\label{h21phi}\\
U_1(\vk)&=& \frac{1}{\sqrt{2}}\left( \begin{array}{cc}
e^{i \omega} & 1\\
e^{-i \omega}& -1
\end{array}\right)\,,\label{u21phi}
\end{eqnarray}
where $\omega=arg(e^{i\phi}+\sin\phi)$. If $H_1 = \tau_\beta H_2\tau_\beta$, then
the conclusions are the same as for model-1 loops, with the replacement
$\tilde p\rightarrow \tilde p|e^{i\phi}+\sin\phi|$.

We now consider the case where the two loops are related through the reflection operation in Eq. (\ref{reflection}).  
Because the reflection implies $\phi\rightarrow -\phi$, the energy dispersions are different for $H_1$ and $H_2$.
In the non-degenerate case ($\delta>0$),  $H^{FS}$ now takes the form
\begin{eqnarray}
H^{FS}=\left( \begin{array}{cc} \tilde p |e^{i\phi}+\sin\phi|-\delta & 
\Delta_{FS}(\vk) \\ 
\Delta_{FS}^*(\vk)
& \tilde p|e^{i\phi}-\sin\phi| -\delta  \end{array}\right)
\end{eqnarray}
and the resulting spectrum allows gapless excitations, as was the case for model-1 loops. 

In the degenerate case, we find it more convenient not to perform the rotation in Eq.  (\ref{UBdGU}), and diagonalize the
original BdG matrix restricted to the subspace 
 $\left(  \hat{\boldsymbol c}_{\vk,1} \,,     \hat{\boldsymbol c}_{-\vk, 2}^\dagger     \right)$, instead. 
In this subspace, the two diagonal blocks of the BdG matrix, which follow from Eq. (\ref{BdGmatrix}), are
$ H_1(\phi)$  and 
  \begin{eqnarray}
  -H_2^T(-\phi)&=&- \tau_\beta H_1^T(\phi)\tau_\beta\,,
  \label{H1H2phi}
\end{eqnarray}
which  follows from (\ref{h21phi}) and  the reflection operation that relates both loops:
  $H_2(\phi)=\tau_\beta H_1(-\phi)\tau_\beta$. 
For $t_\mu=t_1$ (interloop triplet) the BdG reads:
 \begin{eqnarray}
H_{12}'=\left( \begin{array}{cc}
H_1\phi)  & (d_0 + {\boldsymbol d}\cdot{\boldsymbol \tau})i\tau_2  \\
-i\tau_2 (d_0 + {\boldsymbol d}\cdot{\boldsymbol \tau})  & - \tau_\beta H_1^T(\phi)\tau_\beta 
\end{array}\right)
\label{UBU3}
\end{eqnarray}
We identify the TRS cases where the excitation spectrum is fully gapped. 
For constant $d_0$  and  ${\boldsymbol d}=0$, the gapped spectra are obtained for $\beta=1$ and $\beta=3$, respectively:
\begin{eqnarray}
\beta=1:  E^2 &=&d_0^2 + {\tilde p}^2 \sin^2\phi + \left(  d_0\pm \tilde p| \sin\phi+\cos\phi |  \right)^2\,,\nonumber
\\ \\
\beta=3: E^2 &=&d_0^2 + {\tilde p}^2 \left[ \sin^2\phi + (\sin\phi+\cos\phi )^2\right]\,.
\end{eqnarray}
In the case of interloop singlet $t_\mu=t_2$,  we consider $d_0=0$ and constant $\boldsymbol d$. 
Gapped spectra exist for:  $\beta=0$ and nonzero $d_3$; $\beta=1$ and nonzero $d_2$;
$\beta=2$ and nonzero $d_1$. All these cases have similar spectra:
\begin{subequations}
\begin{eqnarray}
\beta=0:  E^2 &=&d_3^2 +  {\tilde p}^2 \left[ \sin^2\phi + (\sin\phi+\cos\phi )^2\right] \,,\ \ \\
\beta=1: E^2 &=&d_2^2 +  {\tilde p}^2 \left[ \sin^2\phi + (\sin\phi+\cos\phi )^2\right]\,,\ \ \\
\beta=2: E^2 &=&d_1^2 +  {\tilde p}^2 \left[ \sin^2\phi + (\sin\phi+\cos\phi )^2\right]\,.\ \
\end{eqnarray}
\end{subequations}

\subsection{Pairing between Dirac loops}

Including  the spin degree of freedom, we may discuss the pairing between spin degenerate loops
$H_1\otimes\sigma_0$ and $H_2\otimes\sigma_0$ described by the BdG matrix:
\begin{eqnarray}
H_{12,s}'=\left( \begin{array}{cc}
H_1\otimes\sigma_0 & (d_0 + {\boldsymbol d}\cdot{\boldsymbol \tau})i\tau_2 (t_\mu)_{12}\sigma_s  \\
-i\sigma_s\tau_2 (d_0 + {\boldsymbol d}\cdot{\boldsymbol \tau})(t_\mu)_{21}  & -H_2^T \otimes\sigma_0 
\end{array}\right)\nonumber\\
\end{eqnarray}
Whatever the choice for $s=1,2,3$, $H_{12,s}'$ decouples in subblocks for which the results obtained above 
for Weyl systems  can be applied.
The (anti)symmetric property of the matrices $t_\mu$, $\sigma_s$ will determine whether the functions
$d_0$,$\boldsymbol d$ should be  odd of even: if for instance, $s=1,3$ then the parity of $d_0$,$\boldsymbol d$
is as in the Weyl case; if, however, $s=2$ (spin singlet), then the parities should be  reversed.

\section{Summary and Conclusions}
 \label{concsec}

We have  described  broken symmetry phases of nested Weyl and Dirac NLs that are induced by a short range interaction. 
We made a systematic analysis for two-band Hamiltonians with PT symmetry, where the
 two nested Weyl NLs can be mapped onto each other through a rotation or
reflection operator. Charge and (pseudo)spin density waves always lower the energy and the broken symmetry phase
can be metallic, semimetallic or insulating, depending on the operator that maps the  
the initial NLs  onto each other, and on whether they enjoy a local reflection symmetry in the loop plane. 
This outcome does not depend
on whether the initial system is semimetallic or metallic (when the initial FS is composed of two toroidal FSs, one hole- and one electron-like). 
If the initial system is semimetallic, spontaneous symmetry breaking requires a finite interaction which is
attractive for  CDWs and repulsive for  PSDWs.

We have also studied specific four-band models, including   the $\mathbb{Z}_2$ NLs,
spin degenerate Dirac systems, and 
 NL's  derived from perturbed spinful Dirac nodal points.
The PSDW phases from $\mathbb{Z}_2$ NLs include  nodal point and  nodal chain semimetals.

Fully gapped superconducting phases from electron pairing in different NLs (interloop pairing), with TRS,  have been found.
They include all possibilities of triplet and singlet pairing in loop space and spin space.

There has recently been an
 intensive search for topological semimetal materials.    
Given that point nodes tend to appear in pairs for symmetry reasons, it is conceivable 
that suitable engineering can produce double NLs. 
Indeed, a recent proposal for realizing point nodes (Dirac or Weyl),  
and  pairs of NLs by strain engineering in
SnTe and GeTe is relevant here\cite{LauOrtix}.  
Another recent proposal concerns layered ferromagnetic rare-earth-metal monohalides 
{\it LnX} ({\it Ln}=La,
Gd; {\it X}=Cl, Br) 
and a pair of mirror-symmetry protected nodal lines in La{\it X}  and Gd{\it X}\cite{Nie}.
Also, splitting of Dirac rings into pairs of Weyl rings by spin-orbit interaction 
in InNbS$_2$ has been proposed\cite{Du}.
Two groups of Dirac nodal rings have been experimentally
detected\cite{Lou} in  ZrB$_2$. 
However, the detection of pairs of NLs at the Fermi level is presently still lacking.

We have not addressed the competition between different orderings or interactions,
but such an extension of our work might be relevant to real materials.  
We have also neglected the effect of the long-ranged tail of the Coulomb interaction
which could be present if the starting system is a NL semimetal with the screening radius
diverging near the Fermi level. In this respect, the study in Ref[\cite{roy}] for a single NL 
suggests that the critical interaction strength 
for orderings where a fully gapped spectrum
arises could be lowered.

Ordered phases\cite{gruner}, such as orbital and/or spin density waves, can be detected through 
neutron scattering, or resonant soft x-ray scattering\cite{xray}. 
The band structure itself may be studied with angle-resolved photoemission spectroscopy.

\section*{Acknowledgments}

M.A.N.A. acknowledges partial support from 
Funda\c{c}\~ao para a Ci\^encia e Tecnologia (Portugal)
through Grant No. UID/CTM/04540/2013, and 
the hospitality of Computational Science Research Center,
 Beijing, China, where this work was initiated.
M.A.N.A. would like to thank  V\'{\i}tor R. Vieira,
  Bruno Mera, and Tilen Cadez for a discussion.

 \vspace{0.5cm}
 
\appendix
\section{Energy spectrum  for Hamiltonian (\ref{rotatedloops})}
\label{appa}

Taking   $\alpha=0$, (CDW case), for instance, and $(a,b)=(1,2)$, 
  the rotated effective Hamiltonian in Eq. (\ref{rotatedloops}) then reads:
 \begin{equation}
A H_{eff}(\vk)  A^\dagger  = 
\left( p_{\parallel} - p_o \right) t_0\tau_1+  p_\perp t_0\tau_2 +U\bar m\ t_1 \tau_{\beta}- \delta\ t_3\tau_0\,.
\label{AHAdelta}
\end{equation}
We perform a suitable rotation on the Hamiltonian $AH_{eff}A^\dagger$ 
in Eq. (\ref{AHAdelta}) so that its energy spectrum can be written
down by inspecting the (anti)commutation relations among its terms.
If $\beta=1$ or 2, we introduce a SU(2) rotation in $t_\mu$ space so that in the end,
only the matrix $t_3$ appears. The required rotation is 
\begin{equation}
W=\cos\frac \theta 2 -it_2\tau_{\beta} \sin\frac \theta 2 \,,
\end{equation}
with the rotation angle $\theta$ given by
\begin{subequations}
\begin{eqnarray}
\sin\theta &=& \frac{U\bar m}{\sqrt{U^2\bar m^2 + \delta^2}}\,,\\
\cos\theta &=& \frac{\delta}{\sqrt{U^2\bar m^2 + \delta^2}}\,,
\end{eqnarray}
\end{subequations}
so that  the rotated Hamiltonian for $\beta=1$ reads
\begin{eqnarray}
WAH_{eff}A^\dagger W^\dagger &=& 
 \left( p_{\parallel} - p_o \right)\tau_1 + p_\perp\cos\theta\ \tau_2 \nonumber\\
  &+& p_\perp\sin\theta\ t_2\tau_3\nonumber\\
 &-&  \sqrt{U^2\bar m^2 + \delta^2}\ t_3\,,
\end{eqnarray}
and the energy spectrum obeys
\begin{eqnarray}
E^2&=& \left[ 
\sqrt{  \left( p_{\parallel} - p_o \right)^2 + p_\perp^2\cos^2\theta}\pm
\sqrt{U^2\bar m^2 + \delta^2}
\right]^2 \nonumber\\
 &+&  p_\perp^2\sin^2\theta\,,
\end{eqnarray}
 which is equivalent to Eq. (\ref{CDWka}).
 
 For $\beta=2$ we have
\begin{eqnarray}
WAH_{eff}A^\dagger W^\dagger &=&
 \left( p_{\parallel} - p_o \right)\cos\theta\ \tau_1  + 
 p_\perp\tau_2 \nonumber\\ &-&  \left( p_{\parallel} - p_o \right)\sin\theta\  t_2\tau_3\nonumber\\
 &-&  \sqrt{U^2\bar m^2 + \delta^2}\ t_3\,,
\end{eqnarray}
and the energy spectrum obeys
\begin{eqnarray}
E^2&=& \left[ 
\sqrt{  \left( p_{\parallel} - p_o \right)^2 \cos^2\theta+ p_\perp^2}\pm
\sqrt{U^2\bar m^2 + \delta^2}
\right]^2 \nonumber\\
&+&  \left( p_{\parallel} - p_o \right)^2\sin^2\theta\,,
\end{eqnarray}
which is equivalent to Eq. (\ref{CDWkb}).

In the case $\beta=3$, it is preferable to perform a SU(2) rotation in $\tau$ space in order 
to eliminate one of the Pauli matrices $\tau$. To this aim, we introduce
\begin{equation}
R_c=\cos\frac \theta 2 -i\tau_3\sin\frac \theta 2 \,,
\end{equation}
with 
\begin{subequations}
\begin{eqnarray}
\sin\theta &=& \frac{ p_{\parallel} - p_o }{\sqrt{\left( p_{\parallel} - p_o \right)^2 + \delta^2}}
\,,\\
\cos\theta &=& \frac{p_\perp}{\sqrt{ \left( p_{\parallel} - p_o \right)^2 + \delta^2}}\,,
\end{eqnarray}
\end{subequations}
so that the rotation of the Hamiltonian now works out as,
\begin{eqnarray}
R_cAH_{eff}A^\dagger R_c^\dagger &=&
{\sqrt{ \left( p_{\parallel} - p_o \right)^2 + \delta^2}}\ \tau_2 + U\bar m \ t_1\tau_3-\delta \ t_3\,.\nonumber\\
\end{eqnarray}

More generally, 
if the product $\tau_\alpha\tau_{\beta}=i\epsilon_{\alpha\beta j}\tau_j$, 
then the above results for the energy still hold, because the appearance of the factor $i$ in the $U\bar m$ term 
would lead to the replacement $t_1\rightarrow t_2$ in  (\ref{AHAdelta}),  
 which does not change the (anti)commutation relations among the Hamiltonian terms.

\section{Mean field treatment of PSDW/CDW}
\label{APMF}

Given the order parameter for a PSDW, 
$ \langle n_s(\br)\rangle  = \frac 1 2 n + \bar m (-1)^s\cos({\vQ}\cdot\br )$,
 one may transform to Fourier space as
\begin{eqnarray}
\langle n_s(\br)\rangle  =\frac 1 N\sum_{\vq} \langle  c_{\vq s}^\dagger c_{\vq s} \rangle +
e^{i\vQ\cdot\br}\langle  c_{\vq s}^\dagger c_{\vq+\vQ s} \rangle + 
e^{-i\vQ\cdot\br}\langle  c_{\vq +\vQ s}^\dagger c_{\vq s} \rangle\,.\nonumber\\
\end{eqnarray}
Using  $\hat\psi_s(\br) =\sum_{\vq} e^{i\vq\cdot\br} c_{\vq s}/\sqrt{N}$, where
 $N$ is the number of momentum values
in the summation,   we see that the above $\langle n_s(\br)\rangle$  is obtained if 
\begin{eqnarray}
\frac 1 N\sum_{\vq} \langle  c_{\vq s}^\dagger c_{\vq s} \rangle&=& \frac 1 2 n\,,\\
\frac 1 N\sum_{\vq} \langle  c_{\vq+\vQ s}^\dagger c_{\vq s} \rangle =
\frac 1 N\sum_{\vq} \langle  c_{\vq  s}^\dagger c_{\vq +\vQ s} \rangle &=&
 \frac 1 2  \bar m (-1)^s\,, \label{MFQ}
\end{eqnarray}
for $\alpha=3$ (PSDW). 
If $\alpha=0$ (CDW) then the factor $(-1)^s$ should be omitted. 
The Hamiltonian is given by
\begin{eqnarray}
\hat H_{eff} = \frac 1 2\sum_{\vk}
(\hat {\boldsymbol  c}_{\bfk}^\dagger \hat {\boldsymbol  c}_{\bfk+\vQ}^\dagger) H_{eff}(\bfk)
  \left( \begin{array}{c}\hat {\boldsymbol  c}_{\vk }\\ \hat {\boldsymbol  c}_{\vk+\vQ } \end{array}  \right)
\end{eqnarray}
where $\hat{\boldsymbol c}_{\vk} = (\hat c_{\vk ,1} \hat c_{\vk ,2})$. We assume that 
$H_{eff}(\bfk)$ is diagonalized by a unitary matrix, $S$,   so that
$SH_{eff}(\bfk)S^\dagger$ is the diagonal matrix composed of the eigenenergies. Then, 
the operators $\hat{\boldsymbol\gamma}$ which destroy the elementary excitations 
 are given by
\begin{eqnarray}
\hat{\boldsymbol\gamma} = S 
 \left( \begin{array}{c}\hat {\boldsymbol  c}_{\vk}\\ \hat {\boldsymbol  c}_{\vk+\vQ } \end{array}  \right)\,.
 \label{gammaS}
\end{eqnarray}
Following Eq. (\ref{MFQ}) we can see that
\begin{eqnarray}
\frac 1 N\sum_{\vk}  \langle
(\hat {\boldsymbol  c}_{\bfk}^\dagger \hat {\boldsymbol  c}_{\bfk+\vQ}^\dagger) 
 \left( \begin{array}{cc} 
 0 & \tau_\alpha \\  \tau_\alpha  & 0 \end{array}  \right)
  \left( \begin{array}{c}\hat {\boldsymbol  c}_{\vk }\\ \hat {\boldsymbol  c}_{\vk+\vQ } \end{array}  \right) 
  \rangle
   = \mp 2\bar m \left\{
   \begin{array}{c} \alpha=3\\ \alpha=0\end{array}
  \right.
\end{eqnarray}
or, in the eigenbasis using (\ref{gammaS}), 
\begin{eqnarray}
&&\frac 1 N\sum_{\vk} \langle \hat{\boldsymbol\gamma}^\dagger S
\left( \begin{array}{cc} 
 0 & \tau_\alpha \\  \tau_\alpha  & 0 \end{array}  \right) S^\dagger\hat{\boldsymbol\gamma} \rangle\ 
 =\mp 2\bar m \left\{
   \begin{array}{c} \alpha=3\\ \alpha=0\end{array}
  \right. \nonumber\\
&=&
\frac 1 N\sum_{\vk} 
\sum_j \left[ 
S
\left( \begin{array}{cc} 
 0 & \tau_\alpha \\  \tau_\alpha  & 0 \end{array}  \right) S^\dagger
\right]_{jj} f(E_j(\vk)) 
   \,,
  \label{SalphaS}
\end{eqnarray}
where $f(x)=1/(1+e^{x/T})$ denotes the Fermi-Dirac distribution function, and $j=1,...,4$ denotes a band index.
The energy dispersions, $E_j(\vk)$, are given in the main text.
However, it is more convenient to work with the transformed Hamiltonian $A H_{eff}A^\dagger$ as in the main text,
which implies that all operators are similarly rotated and $S\rightarrow SA^\dagger$ above.
 Then, Eq. (\ref{SalphaS}) can be written in the form,
\begin{eqnarray}
&&\frac 1 N\sum_{\vk} 
\sum_j \left[ 
SA^\dagger
\left( \begin{array}{cc} 
 0 & \tau_\alpha\tau_{\beta} \\  \tau_{\beta}\tau_\alpha  & 0 \end{array}  \right)
AS^\dagger
\right]_{jj} f(E_j(\vk))\nonumber\\
&=& \mp 2\bar m \left\{
   \begin{array}{c} \alpha=3\\ \alpha=0\end{array}
  \right. \,.\nonumber\\
  \label{APMFeq}
\end{eqnarray}

\section{Critical interaction $U_{cr}$ for degenerate NLs}
\label{MFloop}

We consider a circular nodal line and use the  momentum parametrization in Eq. (\ref{pcoordinates}).

 We linearize the theory in a toroidal region surrounding the NL
up to a momentum cutoff: $0<\tilde p<\tilde p_c$, $0<\theta,\phi<2\pi$.
The volume element is $d^3p=(p_0 + \tilde p\cos\phi)\tilde p\cdot d\tilde p  d\theta d\phi$. 
The number of $\vk$ terms in the toroidal region around the NL is then given by
\begin{eqnarray}
N=\frac{1}{(2\pi\hbar)^3}\int d^3p &=&   \frac{p_0\tilde p_c^2}{4\pi\hbar^3}\,.\label{N}
\end{eqnarray}

In order to simplify the calculations, it is  assumed that 
 the dispersion relation has the same velocity, $v$,  in the NL plane and perpendicular to it.
Then, using $p_\parallel - p_0 = \tilde p\cos\phi$, the model-1 NL Eq. (\ref{loop1}.a)  reads
\begin{eqnarray}
H_0(\vk) = v\tilde p \left( \cos\phi\tau_1 + \sin\phi\tau_2\right)\,,
\end{eqnarray}
where $v=v_1=v_2$ and  we shall take $\delta=0$. As before, we proceed considering the velocity $v=1$ and but shall restore it in the final result
for $U_{cr}$.
In the cases  where $\tau_\alpha \tau_{\beta}\propto\tau_1$ and   $\tau_\alpha\tau_{\beta}\propto\tau_2$
the ordered phases are semimetalic and yield similar mean field equations. 
In the case where $\tau_\alpha\tau_{\beta}\propto\tau_2$, for instance,  the negative energy bands are 
\begin{eqnarray}
E_{1(2)} &=&-\sqrt{ \tilde p^2\pm 2\tilde pU\bar m\sin\phi + U^2\bar m^2}
\end{eqnarray}
and the l.h.s. of Eq. (\ref{APMFeq}) takes the form,
\begin{eqnarray}
&&\frac 1 N\sum_{\vk} 
\sum_j \left[ 
SA^\dagger
\left( \begin{array}{cc} 
 0 & \tau_\alpha\tau_{\beta} \\  \tau_{\beta}\tau_\alpha  & 0 \end{array}  \right)
AS^\dagger
\right]_{jj} f(E_j(\vk)) \nonumber\\ &=&  \frac 1 N\
\int \frac{2(\tilde p\sin\phi-U\bar m) (p_0 + \tilde p\cos\phi)
d^3p }{(2\pi\hbar)^3\sqrt{ \tilde p^2- 2\tilde pU\bar m\sin\phi + U^2\bar m^2}}\,.
\end{eqnarray}
In the limit $U\bar m\rightarrow 0$ one can Taylor expand the integrand to first order. 
The  mean field equations (\ref{APMFeq})   yield
\begin{eqnarray}
\frac{p_0\tilde p_c}{4\pi N\hbar^3} =  \pm \frac{1}{U_{cr}}
\Longrightarrow\
U_{cr}=\pm v \tilde p_c
 \left\{
   \begin{array}{c} \alpha=3\\ \alpha=0\end{array}
  \right. \,,
\end{eqnarray}
where we used (\ref{N}) and  the velocity $v$ has been restored. The finiteness of $U_{cr}$
stems from the linear form of the density of states near the Fermi level. 
The case where $\tau_\alpha\tau_{\beta}\propto\tau_{c\neq a,b}$, where the PSDW 
phase is insulating,
 yields a similar result modified by a prefactor of $1/2$:
$U_{cr}=\pm v \tilde p_c/2$. This is valid also for the case $\tau_\alpha\tau_{\beta}=1$, where the density wave phase is metallic.
We see then that the CDW phase requires an attractive interaction.

\end{document}